\documentclass[10pt,superscriptaddress,nofootinbib,notitlepage,tightenlines,twocolumn, pra]{revtex4-2}

\usepackage{amsthm,amsmath,amssymb}
\usepackage{mathrsfs}
\usepackage{amsthm}
\usepackage{amsmath}
\usepackage{amssymb}
\usepackage{graphicx}
\usepackage{epstopdf}
\usepackage{url}
\usepackage{booktabs}
\usepackage{float}
\usepackage{bm}
\usepackage{multirow}
\usepackage{ifthen}
\usepackage[usenames,dvipsnames]{color}
\usepackage{mathrsfs}
\usepackage[colorlinks=true,citecolor=blue,urlcolor=black]{hyperref}
\usepackage{float}
\usepackage{subfigure}
\usepackage{makecell}
\usepackage{booktabs}
\usepackage{threeparttable}

\newcommand{\be}{\begin{equation}}
\newcommand{\ee}{\end{equation}}

\newcommand{\sket}[1]{{\ensuremath{\lvert#1\rangle}}}
\newcommand{\lket}[1]{{\ensuremath{\left\lvert#1\right\rangle}}}
\newcommand{\ket}[1]{\if@display\lket{#1}\else\sket{#1}\fi}

\usepackage{stackengine}
\newcommand{\sbra}[1]{{\ensuremath{\langle#1\rvert}}}
\newcommand{\lbra}[1]{{\ensuremath{\left\langle#1\right\rvert}}}
\newcommand{\bra}[1]{\if@display\lbra{#1}\else\sbra{#1}\fi}

\newcommand{\sbraket}[2]{{\ensuremath{\langle#1\rvert#2\rangle}}}
\newcommand{\lbraket}[2]{{\ensuremath{\left\langle#1\!\left\rvert\vphantom{#1}#2\right.\!\right\rangle}}}
\newcommand{\braket}[2]{\if@display\lbraket{#1}{#2}\else\sbraket{#1}{#2}\fi}

\newcommand{\sketbra}[2]{{\ensuremath{\lvert #1\rangle\!\langle #2\rvert}}}
\newcommand{\lketbra}[2]{{\ensuremath{\left\lvert #1\right\rangle\!\!\left\langle #2\right\rvert}}}
\newcommand{\ketbra}[2]{\if@display\lketbra{#1}{#2}\else\sketbra{#1}{#2}\fi}

\theoremstyle{plain}

\theoremstyle{definition}

\begin{document}

\title{Asynchronous measurement-device-independent quantum key distribution with hybrid source}

\author{Jun-Lin Bai}
\author{Yuan-Mei Xie}
\affiliation{National Laboratory of Solid State Microstructures and School of Physics, Collaborative Innovation Center of Advanced Microstructures, Nanjing University, Nanjing 210093, China.}
\author{Yao Fu}\email{yfu@iphy.ac.cn}
\affiliation{Beijing National Laboratory for Condensed Matter Physics and Institute of Physics, Chinese Academy of Sciences, Beijing 100190, China.}
\author{Hua-Lei Yin}\email{Corresponding author: hlyin@nju.edu.cn}
\author{Zeng-Bing Chen}\email{zbchen@nju.edu.cn}
\affiliation{National Laboratory of Solid State Microstructures and School of Physics, Collaborative Innovation Center of Advanced Microstructures, Nanjing University, Nanjing 210093, China.}


\begin{abstract}
The linear constraint of secret key rate capacity is overcome by the tiwn-field quantum key distribution (QKD). However, the complex phase-locking and phase-tracking technique requirements throttle the real-life applications of twin-field protocol. The asynchronous measurement-device-independent (AMDI) QKD or called mode-pairing QKD protocol can relax the technical requirements and keep the similar performance of twin-field protocol. Here, we propose an AMDI-QKD protocol with a nonclassical light source by changing the phase-randomized weak coherent state to a phase-randomized coherent-state superposition in the signal state time window. Simulation results show that our proposed hybrid source protocol significantly enhances the key rate of the AMDI-QKD protocol, while exhibiting robustness to imperfect modulation of nonclassical light sources.
\end{abstract}

\maketitle
\section{Introduction}

Quantum key distribution (QKD) can distribute keys between two parties with information-theoretical security, and there has been some progress in security theory and experimental implementation \cite{lo2012measurement,wang2013three, yin2016measurement,  zhou2016making, boaron2018secure,liu2019experimental,wei2020high,liu2021homodyne,wang2022twin,gu2022experimental,zhou2023twin}. Despite no information being revealed from ideal QKD systems, attackers are still able to use the imperfection of real devices to perform attacks on the actual QKD system \cite{lydersen2010hacking, tang2013source, xu2020secure, pirandola2020advances}. Various QKD protocols have been proposed \cite{braunstein2012side, lo2012measurement,lucamarini2018overcoming, xu2020sending, xie2022breaking,zeng2022mode,10.1093/nsr/nwac186,xie2023advantages} and security proof theories \cite{ma2018phase,Wang2018twin, Cui2019Twin, curty2019simple, maeda2019repeaterless, yin2019finite} have been developed to bridge the gap between theoretical and practical conditions. 

The measurement-device-independent (MDI) QKD protocol \cite{lo2012measurement} is able to close all security loopholes of the measurement party, while the transmission distance of MDI-QKD protocols is still relatively close because of two-photon interference in the protocols. For the key transmission distance and key rate of the QKD protocol, there is a theoretical upper limit, the Pirandola-Laurenza-Ottaviani-Banchi (PLOB) bound \cite{pirandola2017fundamental}, which restricts the repeater-less QKD system. It indicates the transmission distance of a single photon in the channel. Luckily, the twin-field (TF) QKD protocol \cite{lucamarini2018overcoming} utilizes the idea of single-photon interference to achieve a breakthrough in repeater-less QKD protocols over long distances while preserving the MDI characteristic of the protocol.

The TF-QKD protocol \cite{lucamarini2018overcoming} has very significant advantages over the MDI-QKD protocol in terms of key rate and transmission distance. However, TF-QKD requires phase tracking and phase locking technology in practical implementation because of the single-photon interference. For realistic scenarios, these techniques significantly increase the difficulty of QKD system implementation. To reduce this difficulty, improved protocols~\cite{xie2022breaking,zeng2022mode} have been proposed by using asynchronous coincidence pairing, which is named as asynchronous measurement-device-independent (AMDI) QKD~\cite{xie2022breaking} or called mode-pairing QKD~\cite{zeng2022mode}. Importantly, both AMDI-QKD experiments~\cite{Zhu2023exp,zhou2022experimental} have been successfully demonstrated in laboratory.
Ref.~\cite{zhou2022experimental} extends the maximal distance from 404 km to 508 km fiber and breaks the repeaterless bound without using phase-tracking and phase-locking. Ref.~\cite{Zhu2023exp} improves the secret key rate by 3 orders of magnitude over 407 km fiber using two independent off-the-shelf lasers and no phase-locking situation.

The data used to form the raw key in the AMDI protocol are derived from the two post-matching time bins of the weak coherent state (WCS) and vacuum state. Furthermore, the security key rate is estimated according to the single photon component of the WCS. The WCS with a random phase can be considered as a photon number mixing state, where multiphoton components are not used to form the raw key in the AMDI-QKD protocol. With the maturity of experimental techniques, the preparation of nonclassical sources in QKD protocol implementation is gradually being considered \cite{yin2014long, zhang2019twin,xu2020hybrid}. Coherent-state superposition (CSS) , also known as cat-state often serves as a basis for quantum cryptography \cite{yin2014long}. There have been some CSS preparation experiments reported \cite{PhysRevLett.97.083604,PhysRevLett.115.023602}. Quantum cryptography tasks require specific CSS amplitudes, and the experiments above are used to prepare CSS with relatively small amplitudes. For the difficulty of controlling amplitudes, a transformation protocol which can iterate to obtain arbitrarily high amplitudes has been proposed \cite{2017Enlargement}. Therefore, QKD protocols with CSS applied have promising application potential in the future. Our work utilizes phase-randomized coherent-state superposition (CSS) in the signal state time window while decoy states time window are still send phase-randomized WCS and significantly improves the performance of the AMDI protocol.

\section{Protocol description}\label{two}

In our hybrid source AMDI protocol, the first two rounds of the process are repeated $N$ times to obtain enough data, and the signal state and the decoy state are post-selected accordingly. To send the states that will be selected as the decoy state, we assume that the time bin subscript is $n\in\left\{1,2,...,N\right\}$, and the phase, bit value and intensity are $\theta_{a(b)}^n\in\left[0,2\pi\right)$, $r_{a(b)}^n\in\left\{0,1\right\}$ and $k_{a}^n\in\left\{\nu_a,\omega_a,o_a\right\},~\nu_a>\omega_a>o_a=0$, or $k_{b}^n\in\left\{\nu_b,\omega_b,o_b\right\},~\nu_b>\omega_b>o_b=0$. Then, Alice and Bob send the prepared states $\left|e^{i(\theta_a^n+r_a^n\pi)}\sqrt{k_a^n}\right \rangle$ and $\left|e^{i(\theta_b^n+r_b^n\pi)}\sqrt{k_b^n}\right \rangle$ to Charlie at the measurement part, respectively. For the time bin of the signal state, the coherent states of the light intensity $\mu_a(\mu_b)$ with the same amplitude and opposite phase form the CSS, $\frac{1}{\sqrt{N_{-,a(b)}}}\left(\left |\alpha_A(\alpha_B) \right \rangle -\left |-\alpha_A(\alpha_B)\right \rangle \right)$, where $N_{-,a(b)}=2(1-e^{-2\mu_a(\mu_b)})$, and $|\alpha_A(\alpha_B)|^2=\mu_a(\mu_b)$. When the phases of CSS from Alice and Bob are randomized, the CSS sent in our proposed hybrid protocol is a mixture of Fock states, and the density matrix of the sent CSS is expanded in the photon number space as $\rho=\sum_{n=0}^{\infty}P_{\xi(\mu)}(2n+1)\left| 2n+1 \right\rangle\left\langle 2n+1\right|$, where $P_{\xi(\mu)}(2n+1)=\frac{\mu^{2n+1}}{\sinh(\mu)(2n+1)!}$. For simplicity, we use $\xi (\mu_a(\mu_b))$ to denote $\frac{1}{\sqrt{N_{-,a(b)}}}\left(\left |\alpha_A(\alpha_B) \right \rangle -\left |-\alpha_A(\alpha_B)\right \rangle \right)$. Alice (Bob) sends CSS $\xi (\mu_a(\mu_b))$ with probability of $p_{\xi (\mu_a(\mu_b))}$, and sends WCS $\nu_{a(b)}$, $\omega_{a(b)}$, $o_{a(b)}$ with probability of $p_{\nu_{a(b)}}$, $ p_{\omega_{a(b)}}$, $p_{o_{a(b)}}$. In the subsequent measurement step, Charlie at the measurement part performs an interference measurement for each time bin (his observations of detector clicks are recorded as events). The detectors respond with gain of $q(k_a,k_b)$, where $k_{a(b)}\in\left\{\xi (\mu_a(\mu_b)),~\nu_{a(b)},~\omega_{a(b)},~o_{a(b)}\right\}$. Charlie announces whether the event is detected and which detector clicks.

We perform click filtering in our protocol, and only $\left\{k_a^{\rm{tot}}=\xi(\mu_a),k_b^{\rm{tot}}=\xi(\mu_b)\right\}$ is used to form the raw key. The overall pairing after matching the two time bins is denoted as $\left\{k_a^{\rm{tot}},k_b^{\rm{tot}}\right\}$, where $k_{a(b)}^{\rm{tot}}=k_{a(b)}^e+k_{a(b)}^l$ and the $e, l$ superscripts represent the previous moment and the next moment after the pairing, respectively. If either Alice or Bob pairs two response events corresponding to two signal states, $k_{a(b)}^{\rm{tot}}=\xi(\mu_{a}(\mu_{b}))^e+\xi(\mu_{a}(\mu_{b}))^l$, the data of this pairing are discarded. Additionally, if either Alice or Bob pairs two events corresponding to a signal state and a decoy state on one side (Alice or Bob), the data of this pairing will be discarded. Alice (Bob) announces the corresponding event of decoy state $\nu_a$ and $\omega_a$ ($\nu_b$ and $\omega_b$), then Bob (Alice) announces the response events that will be filtered. The detailed process description can be found in Supplement 1.

In the event interval $T_c$, the laser signal is sent at a frequency of $F$, and then $N_{T_c}=FT_c$ signals are sent in the $T_c$ time interval. If a detector response is generated in one time bin, the probability of having at least one detector response event in the subsequent $T_c$ time is $q_{T_c}=1-(1-q_{ tot})^{N_{T_c}}$, where $q_{\rm{tot}}=\sum_{k_a,k_b}p_{k_a}p_{k_b}q(k_a,k_b)-p_{\xi(\mu_a)}p_{\nu_b}q(\xi(\mu_a),\nu_b)-p_{\nu_a}p_{\xi(\mu_b)}q(\nu_a,\xi(\mu_b))-p_{\xi(\mu_a)}p_{\omega_b}q(\xi(\mu_a),\omega_b)-p_{\omega_a}p_{\xi(\mu_b)}q(\omega_a,\xi(\mu_b))-p_{\nu_a}p_{\omega_b}q(\nu_a,\omega_b)-p_{\omega_a}p_{\nu_b}q(\omega_a,\nu_b)$ is the probability of having a click event. Therefore, it takes on average $1+1/q_{T_c}$ of valid corresponding events to form a valid pairing. Thus, we define the total number of valid successful pairing events $n_{\rm{tot}}=\frac{Nq_{\rm{tot}}}{1+1/q_{T_c}}$. After the pairing is completed, Alice and Bob first check the number of successful pairings. If it is greater than a threshold, they will discard the part of the data that cannot be paired according to the protocol, or if the number is less than this threshold, they will abandon the protocol. For the partial pairing data used as the raw key, which corresponds to $\{\xi(\mu_a),\xi(\mu_b)\}$, $\{\xi(\mu_a),o_b\}$, $\left\{o_a,\xi(\mu _b)\right\}$, $\left\{o_a,o_b\right\}$ (also defined as the Z basis), Alice has previously randomly matched two time bins as required, assuming that the state of time bin $i$ is $\xi(\mu_a)$ and the intensity of time bin $j$ is $o_a$. If $i<j$, then Alice sets her bit value to 0, and vice versa Alice sets her bit to 1. Alice informs Bob of the number of sequences $i,j$ in the pairing, and in the time bin corresponding to Bob, if Bob obtains $k_b^{\rm{min}\{i,j\}}=\xi(\mu_b),~ k_b^{\rm{max}\{i,j\}}=o_b $ then he sets his bit value to 0, and vice versa. 

The data in the X basis are given by $\left\{\nu_a,\nu_b\right\}$, $\left\{o_a,\nu_b\right\}$, $\left\{\nu_a,o_b\right\}$, $\left\{\omega_a,\omega_b\right\}$, $\left\{o_a, \omega_b\right\}$, $\left\{\omega_a,o_b\right\}$. Alice (Bob) in time bin $i$ is with the global phase of $\varphi_{a(b)}^i:=\theta_{a(b)}^i+\phi_{a(b)}^i$, where $\phi_{a(b)}^i$ is the phase evolution caused by the channel. We obtain the global phase difference, which is $\varphi^i=\varphi_{a}^i-\varphi_{b}^i$. Alice and Bob randomly choose two events that satisfy $k_a^i=k_a^j,~k_b^i=k_b^j$ and $|\varphi^i-\varphi^j|\in\left( -\delta,~\delta\right]$ or $\in\left( \pi-\delta,~\pi+\delta\right]$, where $\delta$ is the phase matching interval.  Further, we utilize discrete phase randomization to replace
continuous phase randomisation as the same as in \cite{lucamarini2018overcoming}. The phase
interval is split into $M$ phase slices, and any interval is denoted as $\delta k = 2\pi \frac{k}{M}$, with $k = \{0, ..., M-1\}$. In our simulation, $M$ takes the value of 16, and thus the effect of discrete phase randomization can be neglected \cite{Cao_2015}. We then match these events as $\left\{k_a^i k_a^j,k_b^i k_b^j\right\}$. By calculating the classical bits $r_a^i\oplus r_a^j$ and $r_b^i\oplus r_b^j$, Alice and Bob obtain the bit values in the X basis. In addition, Bob always needs to flip his bit values in the Z basis, and in the X basis, Bob partially flips his bit values \cite{xie2022breaking,zeng2022mode}. 

After the above process, Alice and Bob could perform post-processing operations to obtain the key:
\begin{equation}
    \begin{aligned}
    R_{\rm{key}}&=\frac{F}{N}\bigg\{\underline{s}_0^z+\underline{s}_{11}^z\left[1-H_2\left(\overline{\phi}_{11}^z\right)\right]-\lambda_{\rm{EC}}\\
    &-\log_2\frac{2}{\epsilon_{\rm{cor}}} -2\log_2\frac{2}{\epsilon'\hat{\epsilon}}-2\log_2\frac{1}{2\epsilon_{\rm{PA}}} \bigg\},
    \end{aligned}
    \label{eq 00}
\end{equation}
where $N$ is the data size. The parameter $\underline{s}_0^z$ is the lower bound of the vacuum state event number. The single-photon pair successful event number ${s}_{11}^{z}$ and the phase error rate $\phi_{11}^{z}$ are estimated statistically because of imperfect single-photon sources. The underline and overline represent the lower and upper bounds on the parameters, respectively \cite{chernoff1952measure,yin2020tight}. The parameter $\lambda_{\rm{EC}}=n^zfH_2(E^z)$ accounts for the amount of information leaked during the error correction, where $f$ is the error correction efficiency and $H_2(x)=-x\log_2x-(1-x)\log_2(1-x)$ is the binary Shannon entropy function. $n^z$ and $E^z$ are the total number of bits and total error rate in the Z basis. $\varepsilon_{\rm cor}$ is the failure probability of the error verification, and $\varepsilon_{\rm PA}$ refers to the failure probability of privacy amplification. $\varepsilon'$ and $\hat{\varepsilon}$ represent the coefficients when using the chain rules of smooth min-entropy and max-entropy, respectively. $\varepsilon_{\rm sec}=2(\varepsilon'+\hat{\varepsilon}+2\varepsilon_e)+\varepsilon_\beta+{\varepsilon_0}+\varepsilon_1+\varepsilon_{\rm PA}$, where
{$\varepsilon_0$}, $\varepsilon_1$, and $\varepsilon_e$ are the failure probabilities for estimating the terms  {$s_{0}^z$}, $s_{11}^z$, and $\phi_{11}^z $, respectively.

Representation of the nonclassical state in the photon number space is accompanied by probability coefficients of a non-Poisson distribution. For the detailed calculation of the CSS state detector gain, taking $\{\xi(\mu_a)^{\rm{tot}}=\xi(\mu_a)^e+o_a^l,~\xi(\mu_b)^{\rm{tot}}=\xi(\mu_b)^e+o_b^l\}$ as an example, it is estimated as follows. In the Z basis, Alice and Bob send a CSS to Charlie at the measurement part. We assume that the density matrix of the nonideal CSS with phase randomization can be expressed as follows:
\begin{equation}
    \begin{aligned}
    \rho&=\frac{a}{\sinh(\mu)}\sum_{n=0}^{\infty}\frac{\mu^{2n+1}}{(2n+1)!}\left| 2n+1 \right\rangle\left\langle 2n+1\right|\\
    &+\frac{1-a}{\cosh(\mu)}\sum_{m=0}^{\infty}\frac{\mu^{2m}}{(2m)!}\left| 2m \right\rangle\left\langle 2m\right|\\
    &=\sum_{i=0}^{\infty}P_{\xi(\mu)}(i)\left| i \right\rangle\left\langle i\right|,
    \end{aligned}
    \label{eq 01}
\end{equation}
where $a$ represents the imperfectness of the CSS. For Alice sending $\left| i \right\rangle$ photon number state and Bob sending $\left| j \right\rangle$ photon number state, after BS evolution, the quantum state at the front of Charlie's detector can be shown as follows:
\begin{equation}
    \begin{aligned}
    \left| \psi \right\rangle_{\rm{in}}&=\left| i \right\rangle_A\left| j \right\rangle_B=\frac{\left(a^{\dagger}_A\right)^i}{\sqrt{i!}}\frac{\left(a^{\dagger}_B\right)^j}{\sqrt{j!}}\left| 0 \right\rangle,\\
    \left| \psi \right\rangle_{\rm{out}}&=\sum_{p=0}^{i+j}\sum_{k=0}^{j}\frac{\left(-1\right)^{j-k}C_i^{p-k}C_j^k}{\sqrt{2^{i+j}i!j!}}\left(a^{\dagger}_L\right)^p\left(a^{\dagger}_R\right)^{i+j-p}\left| 0 \right\rangle.
    \end{aligned}
    \label{eq 02}
\end{equation}
In terms of $D_L^p$ and $D_R^p$, the corresponding probabilities of the detector clicks are $D_L^p=1-(1-p_d)(1-\eta)^{p},~D_R^p=1-(1-p_d)(1-\eta)^{i+j-p}$. The parameter $\eta$ is estimated by $\eta=\eta_d10^ {-\frac{\alpha l}{20}}$, where $\eta_d$, $\alpha$, and $l$ are the detector efficiency, attenuation coefficient of the fiber, and total transmission distance, respectively. In photon number space, the yield for only the left detector response or only the right detector response are $Y^L_{i,j}=\sum_{p=0}^{i+j}D_L^p\left(1-D_R^p\right)~P_{i,j},~Y^R_{i,j}=\sum_{p=0}^{i+j}\left(1-D_L^p\right)D_R^p~P_{i,j}$, where $P_{i,j}=\left|\sum_{k=0}^{j}\frac{(-1)^{j-k}C_i^{p-k}C_j^k}{\sqrt{2^{i+j}i!j!}}\sqrt{(p)!}\sqrt{(i+j-p)!}\right|^2$. Thus, we can obtain the gain as follows:
\begin{equation}
    \begin{aligned}
    q{\left(\xi(\mu_a),\xi(\mu_b)\right)}=\sum_{i,j=0}^{\infty}P_{\xi(\mu_a)}(i)P_{\xi(\mu_b)}(j)\left(Y^L_{i,j}+Y^R_{i,j}\right),
    \end{aligned}
    \label{eq 03}
\end{equation}
where we have $P_{\xi(\mu)}(2n+1)=\frac{a}{\sinh(\mu)}\frac{\mu^{2n+1}}{(2n+1)!}$, and we have $P_{\xi(\mu)}(2m)=\frac{1-a}{\cosh(\mu)}\frac{\mu^{2m}}{(2m)!}$ .

\section{Performance and Discussion}\label{three}
\label{sec_discussion}

\begin{table}[t]
\centering
\caption{\bf Simulation parameters. $\eta_{d}$ and $p_{d}$ are the detector efficiency and the dark count rate, respectively. $\alpha$ ($dB/km$) and $f$ are the attenuation coefficient of the fiber and error-correction efficiency, respectively. $\epsilon$ is the failure probability parameter. We set the failure parameters $\varepsilon'$, $\hat{\varepsilon}$,
$\varepsilon_e$, $\varepsilon_{\beta}$ , and $\varepsilon_{PA}$ to be the same $\epsilon$.}
\setlength{\tabcolsep}{4mm}
\begin{tabular}{ccccc}
\hline
$\eta_d$ & $p_d$ & $\alpha$ & $f$ & $\epsilon$\\
\hline
$80\%$ & $10^{-8}$ & $0.16 $ & $1.1$ & $10^{-10}$\\

\hline
\end{tabular}
  \label{tab 02}
\end{table}

We list the parameters used for simulation of the hybrid source AMDI protocol in Table \ref{tab 02}. The detailed calculation procedure and the optimized parameter values at typical distance points can be found in Supplement 1, and the absolute PLOB bound formula that we utilized in the simulation result figures is $R=-\log_2(1-10^{-0.16\frac{L}{10}})$.

\begin{figure}[b]
\centering
\includegraphics[width=\linewidth]{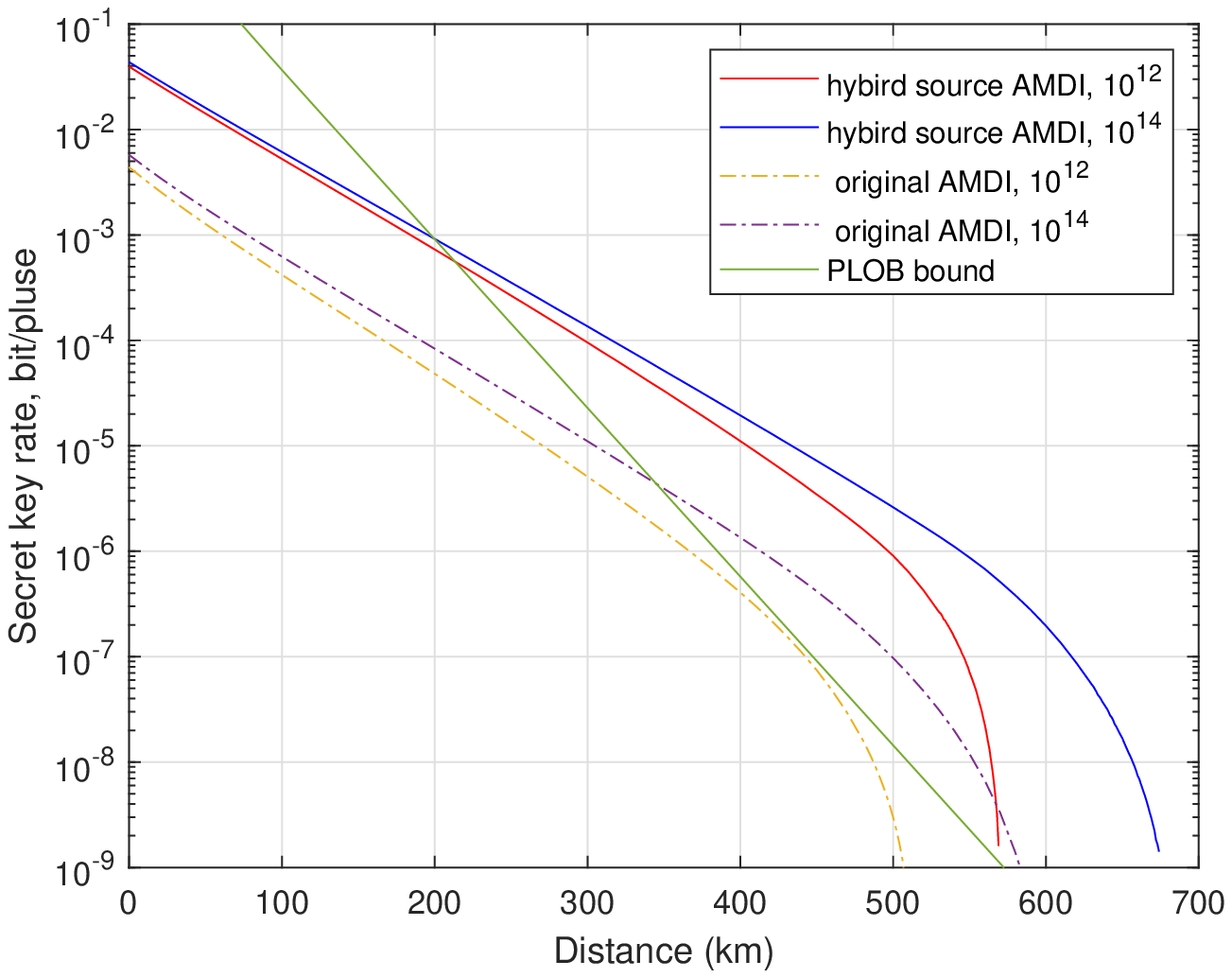}
\caption{Key rate comparison between the perfect CSS AMDI protocol and the original AMDI protocol. The time windows are set as $N=10^{12}$ and $N=10^{14}$.}
\label{N=1214}
\end{figure}

For the perfect CSS state, the parameter $a$ in Eq. \ref{eq 01} takes the value of $1$. Fig. \ref{N=1214} shows the key rate simulation results of the perfect hybrid source AMDI protocol for time windows of $10^{12}$ (corresponding to the red line) and $10^{14}$  (corresponding to the blue line) compared to those of the original AMDI protocol \cite{xie2022breaking,zeng2022mode} for time windows of $10^{12}$ (corresponding to the yellow dotted line) and $10^{14}$  (corresponding to the purple dotted line). The percentage of single photon component of CSS is higher, leading to a higher key rate.

\begin{figure}[t]
\centering
\includegraphics[width=\linewidth]{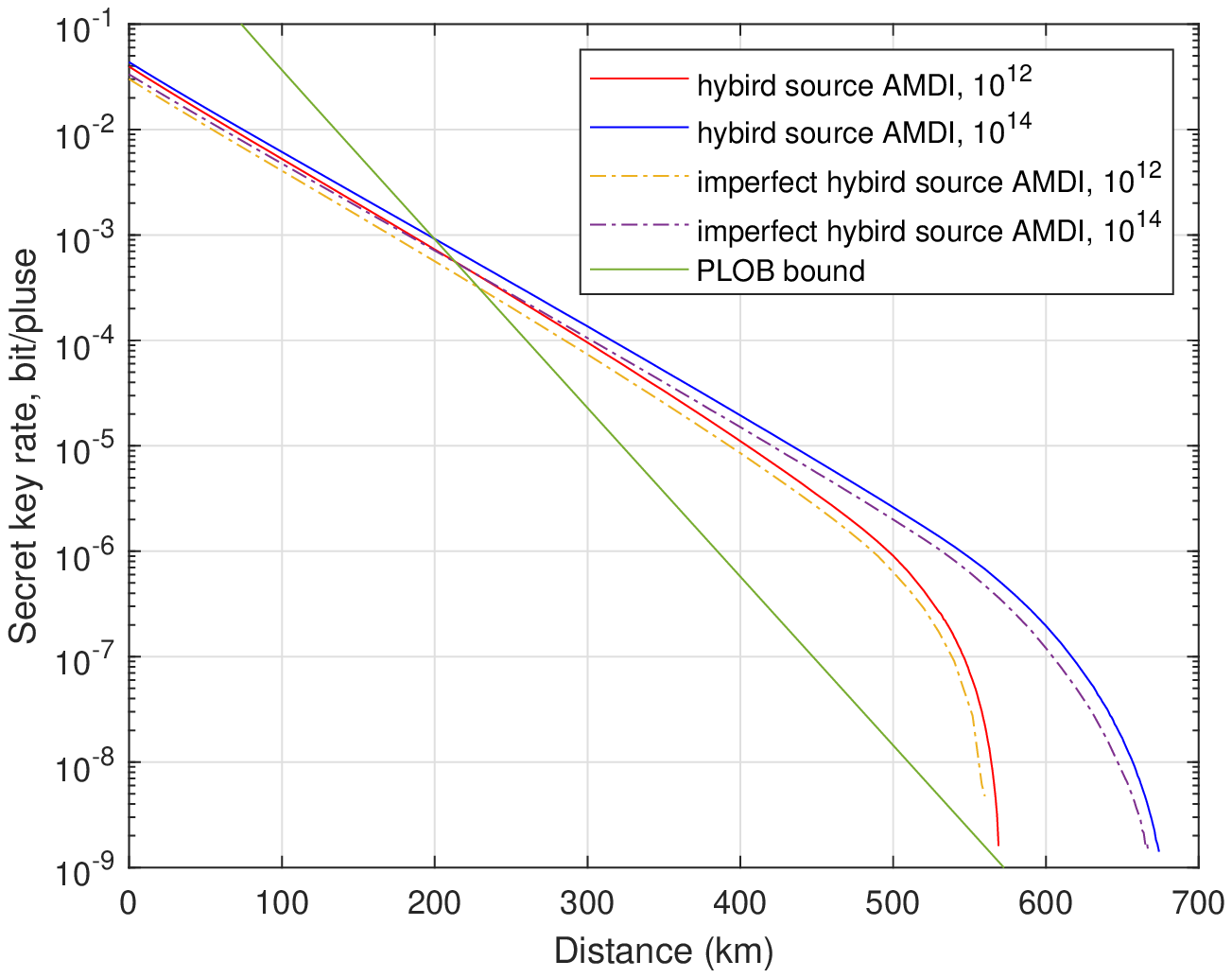}
\caption{Key rate comparison between the perfect CSS AMDI protocol and the imperfect CSS AMDI protocol. The time windows are set as $N=10^{12}$ and $N=10^{14}$.}
\label{N=1214_r}
\end{figure}

Fig. \ref{N=1214_r} shows the performance of the proposed CSS state protocol for time windows of $N=10^{12}$ and $N=10^{14}$ when the signal state is imperfectly modulated. We use the parameter $a$ in Eq. \ref{eq 01} to denote the CSS imperfect modulation of the protocol, and $a$ takes the value of 0.7 in the imperfect CSS AMDI simulation. The results in Fig. \ref{N=1214_r} show that the CSS AMDI protocol is robust to CSS imperfect modulation. The robustness reduces the difficulty of the experimental implementation of the proposed protocol.

We propose an improved AMDI protocol based on nonclassical quantum states. The protocol improves secret key rate by using a quantum state with a higher proportion of single photon components instead of the commonly used WCS source, thus significantly improving the key rate and transmission distance of the AMDI protocol. Our simulations use the CSS as a nonclassical light source. In the event windows of $N=10^{12}$ and $N=10^{14}$, the simulation results of Fig. \ref{N=1214} show that the key rates of the proposed protocol are $1343\%$ and $2624\%$ higher at the distance of 400 km, and the transmission distance is $15.57\%$ and $19.45\%$ farther than the original AMDI protocol, respectively. Besides, the AMDI protocol using the CSS as a nonclassical light source shows that it is robust to imperfect modulation of the nonclassical source, a property that reduces the difficulty of the experimental implementation of the proposed protocol. We remarked that the hybrid source protocol can also be used for other quantum cryptography tasks, for example quantum digital signatures~\cite{yin2023experimental}.

\par
\begin{center}
	\textbf{ACKNOWLEDGMENTS}
\end{center}

This study was supported by  National Natural Science Foundation of China (12274223), Natural Science Foundation of Jiangsu Province (BK20211145), Fundamental Research Funds for the Central Universities (020414380182), Key Research and Development Program of Nanjing Jiangbei New Area (ZDYD20210101),  Program for Innovative Talents and Entrepreneurs in Jiangsu (JSSCRC2021484), and Program of Song Shan Laboratory (included in the management of the Major  Science and Technology Program of Henan Province) (221100210800-02).

\appendix

\section{Pairing method and raw key sifting}\label{supp5}
    \begin{enumerate}
		\item The detailed pairing method is as shown in Fig. \ref{supp_1}. The yellow zones represent the paired data used for decoy estimation. The orange zones stand for $X$ basis quantum bit error rate ($X$-QBER) and decoy estimation. The blue zone is the data prepared for raw key sifting, and the dark grey zones are the discard pairing results.
        \begin{figure}[b]
            \centering
            \includegraphics[width=\linewidth]{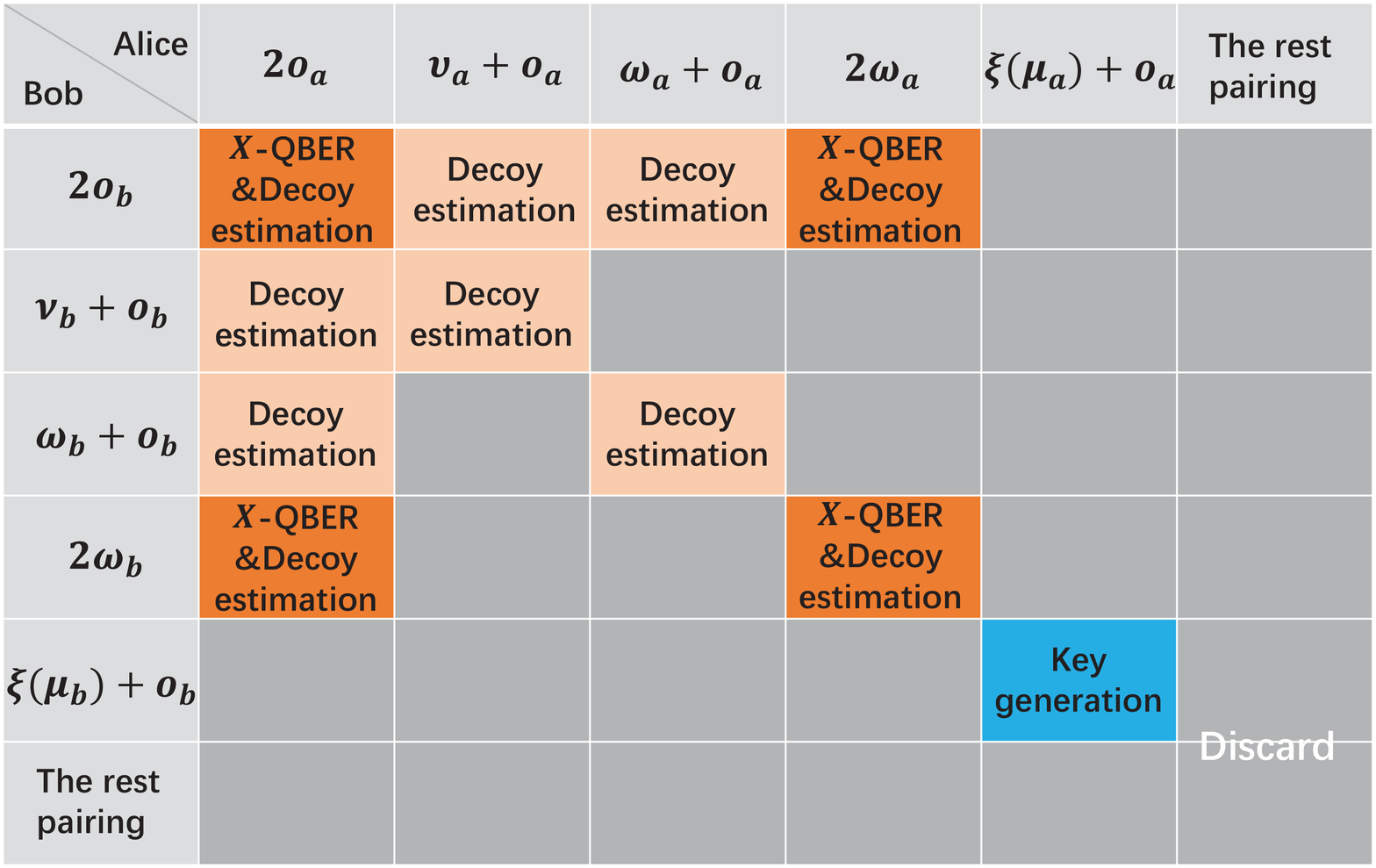}
            \caption{Detailed pairing method of our protocol. The yellow zones represent the paired data used for decoy estimation. The orange zones stand for $X$ basis quantum bit error rate ($X$-QBER) and decoy estimation. The blue zone is the data prepared for raw key sifting, and the dark grey zones are the discard pairing results.}
            \label{supp_1}
        \end{figure}
            \item In the X basis, the phase, bit value and intensity are $\theta_{a(b)}^n\in\left(0,2\pi\right]$, $r_{a(b)}^n\in\left\{0,1\right\}$ and $k_{a}^n\in\left\{\nu_a,\omega_a,o_a\right\},~\nu_a>\omega_a>o_a=0$, or $k_{b}^n\in\left\{\nu_b,\omega_b,o_b\right\},~\nu_b>\omega_b>o_b=0$. Then, Alice and Bob send the prepared states $\left|e^{i(\theta_a^n+r_a^n\pi)}\sqrt{k_a^n}\right \rangle$ and $\left|e^{i(\theta_b^n+r_b^n\pi)}\sqrt{k_b^n}\right \rangle$ to Charlie at the measurement part, respectively. The data in the X basis are given by $\left\{\nu_a,\nu_b\right\}$, $\left\{o_a,\nu_b\right\}$, $\left\{\nu_a,o_b\right\}$, $\left\{\omega_a,\omega_b\right\}$, $\left\{o_a, \omega_b\right\}$, $\left\{\omega_a,o_b\right\}$. Alice (Bob) in time bin $e$ is with the global phase of $\varphi_{a(b)}^e:=\theta_{a(b)}^e+\phi_{a(b)}^e$, where $\phi_{a(b)}^e$ is the phase evolution caused by the channel. We obtain the global phase difference, which is $\varphi^e=\varphi_{a}^e-\varphi_{b}^e$. Alice and Bob randomly choose two events that satisfy $k_a^e=k_a^l$, $k_b^e=k_b^l$ and $|\varphi^e-\varphi^l|\in\left( -\delta,~\delta\right]$ or $\in\left( \pi-\delta,~\pi+\delta\right]$, where $\delta$ is the phase matching interval. Key mapping method in the X basis is as shown in Fig. \ref{supp_2}. 
        \begin{figure}[t]
            \centering
            \includegraphics[width=\linewidth]{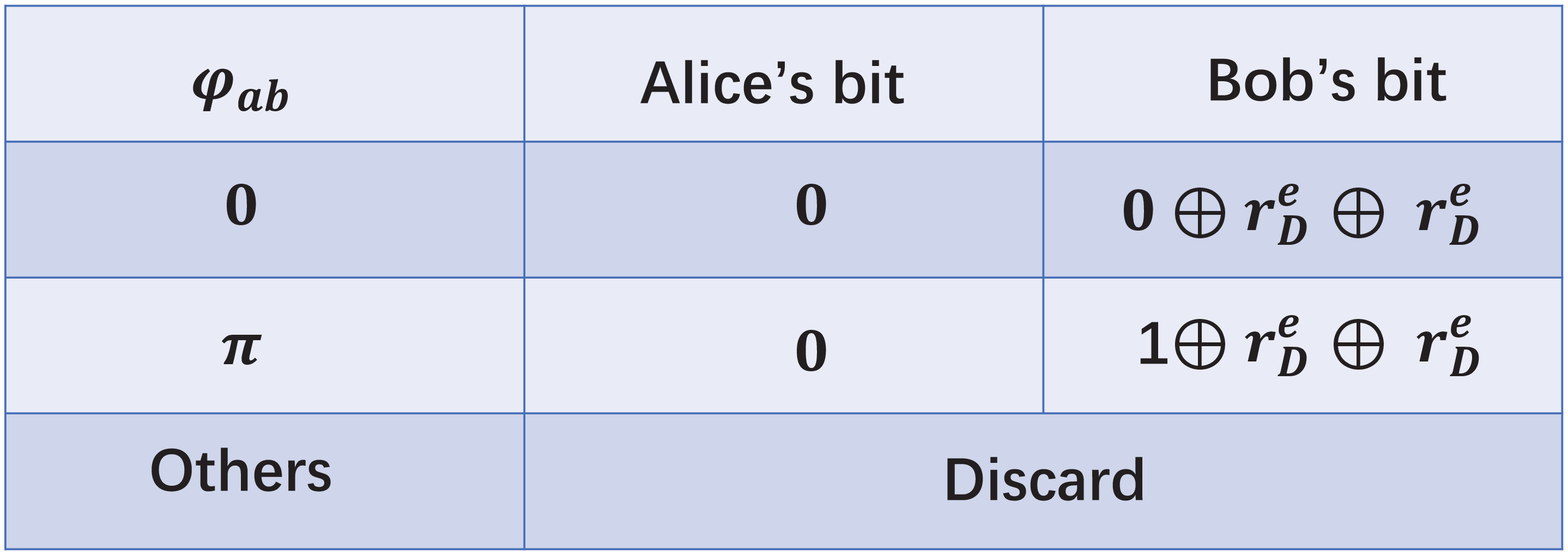}
            \caption{Key mapping in the X basis. Alice (Bob) in earlier time bin $e$ is with the global phase of $\varphi_{a(b)}^e:=\theta_{a(b)}^e+\phi_{a(b)}^e$, where $\phi_{a(b)}^e$ is the phase evolution caused by the channel. We obtain the global phase difference, which is $\varphi^e=\varphi_{a}^e-\varphi_{b}^e$. Alice and Bob randomly choose two events that satisfy $k_a^e=k_a^l,~k_b^e=k_b^l$ and $\varphi_{ab}=|\varphi^e-\varphi^l|=0$ or $=\pi$, where $l$ represents later time bin. In experiment, $=0$ or $=\pi$ are replaced by $\in\left( -\delta,~\delta\right]$ or $\in\left( \pi-\delta,~\pi+\delta\right]$, where $\delta$ is the phase matching interval.}
            \label{supp_2}
        \end{figure}
    \end{enumerate}

\section{Parameter estimation}\label{supp6}

    \subsection{Number of Z-basis single photon pairs}	
	Because of the click filtering process that we describe in the main text, all the intensities of the decoy states are published and only $\left[\xi\left(\mu_a\right),\xi\left(\mu_b\right)\right]$ is used for final key (the middle bracket represents the summation representation of the states of the two time bins $i$, $j$). For simplicity, we let $\nu_a=\nu_b=\nu,~\omega_a=\omega_b=\omega,~o_a=o_b=o$ in the symmetric channel case we consider. The final key of the QKD participants can be extracted from the single-photon pair component finally. With the decoy state method, click filtering, and joint constraints, the lower bound of the Z basis single photon pair component can be estimated as follows.
        \begin{widetext}
        \begin{equation}
		      \begin{aligned}
	           \underline{s}_{11}^{z*}&\geq\frac{p_{\left(1,~\xi\left(\mu_a\right)\right)}p_{\left(1,~\xi\left(\mu_b\right)\right)}p_{\left[\xi\left(\mu_a\right),\xi\left(\mu_b\right)\right]}}{\nu^2\omega^2\left(\nu-\omega\right)} \left[\nu^3 \left(e^{2\omega}\frac{\underline{n}_{\left[\omega,\omega\right]}^{*}}{p_{\left[\omega,\omega\right]}}-e^{\omega}\frac{\overline{n}_{\left[o,\omega\right]}^{*}}{p_{\left[o,\omega\right]}}-e^{\omega}\frac{\overline{n}_{\left[\omega,o\right]}^{*}}{p_{\left[\omega,o\right]}}+\frac{\underline{n}_{\left[o,o\right]}^{*}}{p_{\left[o,o\right]}}\right)\right.\\
				&  \left.-\omega^3 \left(e^{2\nu}\frac{\overline{n}_{\left[\nu,\nu\right]}^{*}}{p_{\left[\nu,\nu\right]}}-e^{\nu}\frac{\underline{n}_{\left[o,\nu\right]}^{*}}{p_{\left[o,\nu\right]}}-e^{\nu}\frac{\underline{n}_{\left[\nu,o\right]}^{*}}{p_{\left[\nu,o\right]}}+\frac{\underline{n}_{\left[o,o\right]}^{*}}{p_{\left[o,o\right]}}\right) \right]\\
                &=\frac{p_{\left(1,~\xi\left(\mu_a\right)\right)}p_{\left(1,~\xi\left(\mu_b\right)\right)}p_{\left[\xi\left(\mu_a\right),\xi\left(\mu_b\right)\right]}}{\nu^2\omega^2\left(\nu-\omega\right)}\left(\underline{S}^*_1-\overline{S}^*_2\right),\label{eq_decoy_s11z}
		      \end{aligned}
        \end{equation}
		where
        
        \begin{equation}
		      \begin{aligned}
                S_1&=\nu^3 e^{2\omega}\frac{\underline{n}_{\left[\omega,\omega\right]}^{*}}{p_{\left[\omega,\omega\right]}}+\omega^3 e^{\nu}\frac{\underline{n}_{\left[o,\nu\right]}^{*}}{p_{\left[o,\nu\right]}}+ \omega^3 e^{\nu}\frac{\underline{n}_{\left[\nu,o\right]}^{*}}{p_{\left[\nu,o\right]}}+(\nu^3-\omega^3 )\frac{\underline{n}_{\left[o,o\right]}^{*}}{p_{\left[o,o\right]}},\\
                S_2&=\omega^3 e^{2\nu}\frac{\overline{n}_{\left[\nu,\nu\right]}^{*}}{p_{\left[\nu,\nu\right]}}+\nu^3 e^{\omega}\frac{\overline{n}_{\left[o,\omega\right]}^{*}}{p_{\left[o,\omega\right]}}+\nu^3e^{\omega}\frac{\overline{n}_{\left[\omega,o\right]}^{*}}{p_{\left[\omega,o\right]}},
	           \label{S1}
		      \end{aligned}
        \end{equation}
        and
        \begin{equation}
            p_{\left(1,~\xi\left(\mu_a\right)\right)}=\frac{4}{N_{-,a}}\mu_ae^{-\mu_a},~p_{\left(1,~\xi\left(\mu_b\right)\right)}=\frac{4}{N_{-,b}}\mu_be^{-\mu_b},~p_{\left[k_a^{\rm{tot}},k_b^{\rm{tot}}\right]}=\sum_{k_a^e+k_a^l}\sum_{k_b^e+k_b^l}\left(\frac{p^e_{k_a}p^e_{k_b}}{p_s}\frac{p^l_{k_a}p^l_{k_b}}{p_s}\right).
        \end{equation}
        
        An asterisk indicates that the variable takes the expected value \cite{chernoff1952measure,yin2020tight}. We use underline and overline to refer to the lower and upper bounds of a variable. When using the click filtering scheme, $p_s=1-p_{\xi\left(\mu_a\right)}p_{\nu_b}-p_{\nu_a}p_{\xi\left(\mu_b\right)}-p_{\xi\left(\mu_a\right)}p_{\omega_b}-p_{\omega_a}p_{\xi\left(\mu_b\right)}-p_{\nu_a}p_{\omega_b}-p_{\omega_a}p_{\nu_b}$. With the joint constraints \cite{yu2015statistical}, $\underline{S}^*_1$, $\overline{S}^*_2$ are the results obtained by applying the statistical fluctuation formula to the mathematical transformation of $S_1$, $S_2$.
	\end{widetext}
 
\subsection{Number of Z-basis vacuum state}
        With click filtering, the Z-basis vacuum state components are counted as
        \begin{equation}
            \underline{s_0^{z*}}=\frac{p_{\left(0,~\xi\left(\mu_a\right)\right)}p_{\left[\xi\left(\mu_a\right),\xi\left(\mu_b\right)\right]}}{p_{\left[o_a,\xi\left(\mu_b\right)\right]}}\underline{n^*_{\left[o_a,\xi\left(\mu_b\right)\right]}}.
        \end{equation}
		For a perfectly modulated coherent superposition state, $p_{\left(0,~\xi\left(\mu_a\right)\right)}$ is equal to zero.

\subsection{Number of X-basis single photon pairs}
	 For simplicity, we let $\nu_a=\nu_b=\nu,~\omega_a=\omega_b=\omega,~o_a=o_b=o$ in the symmetric channel case we consider. With click filtering and joint constraints, the number of successful event pairs for X-basis single photon pairs is as follows, the spatial probability distribution of photon numbers at different light intensities is the same, so we get:
    \begin{equation}
        \begin{aligned}
			\underline{s}_{11}^{x*}\geq&\frac{\omega^2 e^{-4\omega}4p_{\left[2\omega,2\omega\right]}}{\omega^2\nu^2\left(\nu-\omega\right)}\\
            &\times\left[\nu^3 \left(e^{2\omega}\frac{\underline{n}_{\left[\omega,\omega\right]}^{*}}{p_{\left[\omega,\omega\right]}}-e^{\omega}\frac{\overline{n}_{\left[o,\omega\right]}^{*}}{p_{\left[o,\omega\right]}}-e^{\omega}\frac{\overline{n}_{\left[\omega,o\right]}^{*}}{p_{\left[\omega,o\right]}}+\frac{\underline{n}_{\left[o,o\right]}^{*}}{p_{\left[o,o\right]}}\right)\right.\\
				&  \left.-\omega^3 \left(e^{2\nu}\frac{\overline{n}_{\left[\nu,\nu\right]}^{*}}{p_{\left[\nu,\nu\right]}}-e^{\nu}\frac{\underline{n}_{\left[o,\nu\right]}^{*}}{p_{\left[o,\nu\right]}}-e^{\nu}\frac{\underline{n}_{\left[\nu,o\right]}^{*}}{p_{\left[\nu,o\right]}}+\frac{\underline{n}_{\left[o,o\right]}^{*}}{p_{\left[o,o\right]}}\right) \right]\\
                &=\frac{\omega^2 e^{-4\omega}4p_{\left[2\omega,2\omega\right]}}{\omega^2\nu^2\left(\nu-\omega\right)}\left(\underline{S}^*_1-\overline{S}^*_2\right).\label{eq_decoy_s11z0}
		\end{aligned}
  \end{equation}
	where
    
        \begin{equation}
		      \begin{aligned}
                S_1=&\nu^3 e^{2\omega}\frac{\underline{n}_{\left[\omega,\omega\right]}^{*}}{p_{\left[\omega,\omega\right]}}+\omega^3 e^{\nu}\frac{\underline{n}_{\left[o,\nu\right]}^{*}}{p_{\left[o,\nu\right]}}+ \omega^3 e^{\nu}\frac{\underline{n}_{\left[\nu,o\right]}^{*}}{p_{\left[\nu,o\right]}}\\
                &+(\nu^3-\omega^3 )\frac{\underline{n}_{\left[o,o\right]}^{*}}{p_{\left[o,o\right]}},\\
                S_2=&\omega^3 e^{2\nu}\frac{\overline{n}_{\left[\nu,\nu\right]}^{*}}{p_{\left[\nu,\nu\right]}}+\nu^3 e^{\omega}\frac{\overline{n}_{\left[o,\omega\right]}^{*}}{p_{\left[o,\omega\right]}}+\nu^3e^{\omega}\frac{\overline{n}_{\left[\omega,o\right]}^{*}}{p_{\left[\omega,o\right]}}.
	           \label{S2}
		      \end{aligned}
        \end{equation}
    Similarly, the probabilities are calculated as 
    \begin{equation}
        p_{\left[k_a^{\rm{tot}},k_b^{\rm{tot}}\right]}=\sum_{k_a^e+k_a^l}\sum_{k_b^e+k_b^l}\left(\frac{p^e_{k_a}p^e_{k_b}}{p_s}\frac{p^l_{k_a}p^l_{k_b}}{p_s}\right),
    \end{equation}
    apart from $p_{[2\omega_a,2\omega_b]}$ because of the phase matching condition in the $\boldsymbol{X}$ basis, which is
    \begin{equation}
	   \begin{aligned}
	   p_{[2\omega_a,2\omega_b]}=\frac{2}{M}\frac{p_{\omega_a}p_{\omega_b}}{p_{s}} \frac{p_{\omega_a}p_{\omega_b}}{p_{s}}.
	   \end{aligned} 
    \end{equation}
    When using the click filtering scheme, we have $p_s=1-p_{\xi\left(\mu_a\right)}p_{\nu_b}-p_{\nu_a}p_{\xi\left(\mu_b\right)}-p_{\xi\left(\mu_a\right)}p_{\omega_b}-p_{\omega_a}p_{\xi\left(\mu_b\right)}-p_{\nu_a}p_{\omega_b}-p_{\omega_a}p_{\nu_b}$. The parameters $\underline S_1^*$, $\overline S_2^*$ are estimated as in Z basis.

\subsection{Wrong single photon pair pairing event count}
	\begin{enumerate}
		\item For simplicity, we let $\omega_a=\omega_b=\omega,~o_a=o_b=o$ in the symmetric channel case we consider. Wrong single photon pair pairing event number is calculated as following.
        \begin{equation}
        \overline{t}^x_{11}\le m_{\left[2\omega,2\omega\right]}-\underline{m}^0_{\left[2\omega,2\omega\right]},
        \end{equation}
        where $m_{\left[2\omega,2\omega\right]}$ is the observed value of the X basis error pairings number, given in the previous section; $\underline{m}^0_{\left[2\omega,2\omega\right]}$ is the number of error pairing events sent by at least one of Alice or Bob that are vacuum states.
		\item
        The expected lower bound of parameter $\underline{m}^0_{\left[2\omega,2\omega\right]}$ is presented as 
		\begin{equation}
			\begin{aligned}
				\underline{m}^{0*}_{\left[2\omega,2\omega\right]}=&e^{-2\omega}\frac{p_{\left[2\omega,2\omega\right]}}{2p_{\left[o,2\omega\right]}}\underline{n}^*_{\left[o,2\omega\right]}+e^{-2\omega}\frac{p_{\left[2\omega,2\omega\right]}}{2p_{\left[2\omega,o\right]}}\underline{n}^*_{\left[2\omega,o\right]}\\
                &+e^{-2\omega-2\omega}\frac{p_{\left[2\omega,2\omega\right]}}{2p_{\left[o,o\right]}}\underline{n}^*_{\left[o,o\right]}.
			\end{aligned}
		\end{equation}

	\end{enumerate}
	
\subsection{Phase error rate}
    Using $\underline{s}_{11}^x$, the value of the phase error rate $\overline{\phi}_{11}^{z}$ is obtained using the fact that phase error rate in the Z basis is equal to bit error rate in the X basis \cite{chernoff1952measure,yin2020tight}:
	\begin{equation}
		\begin{aligned}
			\overline{\phi}_{11}^{z}=& \overline{e}_{11}^x=\frac{\overline{t}_{11}^x}{\underline{s}^x_{11}}.	
		\end{aligned}\label{eq_asynchronous-MDIQKD_phi11z}
	\end{equation}

\section{DETAILED THEORETICAL MODEL AND SIMULATION}
        When Alice and Bob send coherent state pulses of intensity $k_a$ and $k_b$ (with a phase difference of $\theta$) in X basis, the probability corresponding to that only the left detector (L) or only the right detector (R) responds is
     
        \begin{equation}
            \begin{aligned}
            q_{\left(k_a,k_b\right)}^{\theta,L}&=\left(1-p_d\right)e^{-\frac{\left(\eta k_a+\eta k_b\right)}{2}}e^{\eta\sqrt{ k_a k_b}\cos\theta}\\
            &\times\left[1-\left(1-p_d\right)e^{-\frac{\left(\eta k_a+\eta k_b\right)}{2}}e^{-\eta\sqrt{ k_a k_b}\cos\theta}\right]
            \end{aligned}
        \end{equation}

        or
        \begin{equation}
            \begin{aligned}
            q_{\left(k_a,k_b\right)}^{\theta,R}&=\left(1-p_d\right)e^{-\frac{\left(\eta k_a+\eta k_b\right)}{2}}e^{-\eta\sqrt{ k_a k_b}\cos\theta}\\
            &\times\left[1-\left(1-p_d\right)e^{-\frac{\left(\eta k_a+\eta k_b\right)}{2}}e^{\eta\sqrt{ k_a k_b}\cos\theta}\right],
            \end{aligned}
        \end{equation}
        where $p_d$ is the dark count rate of the detector $D_L$(and $D_R$) for the symmetric channel distance. The parameter $\eta$ stands for $\eta=\eta_d10^ {-\frac{\alpha l}{20}}$, where $l$ is the total transmission distance and $\eta_d$ is the detector efficiency. By integrating the angle $\theta$ from 0 to $2\pi$, the probability that at least one detector will respond when Alice and Bob send pulses with light intensities $k_a$ and $k_b$ is obtained as follows.
        \begin{equation}
            \begin{aligned}
                q{\left(k_a,k_b\right)}&=\frac{1}{2\pi}\int\limits_{0}^{2\pi}\left(q_{\left(k_a,k_b\right)}^{\theta,L}+q_{\left(k_a,k_b\right)}^{\theta,R}\right)d\theta\\
                &=2\left(1-p_d\right)e^{-\frac{\left(\eta k_a+\eta k_b\right)}{2}}I_0\left(\eta\sqrt{ k_a k_b}\right)\\
                &-2\left(1-p_d\right)^2e^{-{\left(\eta k_a+\eta k_b\right)}},
            \end{aligned}
        \end{equation}
     
        where $I_0\left(x\right)$ is a zero-order first kind modified Bessel function.

    \begin{table*}[t]
    \centering
    \caption{\bf The optimized parameter values at typical distance points.}
    \setlength{\tabcolsep}{6mm}
    \begin{tabular}{cccccccc}
    \hline
    Distance (km) & $\mu_a$ & $\nu_a$ & $\omega_a$ & $p_{\xi(\mu_a)}$& $p_{\nu_a}$ & $p_{\omega_a}$ & $T_c$ ($\mu$s)\\
    \hline
    0 & $0.088$ & $0.087$ & $0.036$ & $0.333$& $0.007 $ & $0.038$ & $0.315$\\
    \hline
    200 & $0.104$ & $0.103$ & $0.039$ & $0.276$& $0.015 $ & $0.093$ & $0.397$\\
    \hline
    400 & $0.180$ & $0.179$ & $0.051$ & $0.225$& $0.026 $ & $0.209$ & $0.425$\\

    \hline
    \end{tabular}
    \label{tab 01}
    \end{table*}

		As we define in the main text, the total number of valid successful pairing events can be calculated as $n_{\rm{tot}}=\frac{Nq_{\rm{tot}}}{1+1/q_{T_c}}$, where $q_{\rm{tot}}=\sum_{k_a,k_b}p_{k_a}p_{k_b}q(k_a,k_b)-p_{\xi(\mu_a)}p_{\nu_b}q(\xi(\mu_a),\nu_b)-p_{\nu_a}p_{\xi(\mu_b)}q(\nu_a,\xi(\mu_b))-p_{\xi(\mu_a)}p_{\omega_b}q(\xi(\mu_a),\omega_b)-p_{\omega_a}p_{\xi(\mu_b)}q(\omega_a,\xi(\mu_b))-p_{\nu_a}p_{\omega_b}q(\nu_a,\omega_b)-p_{\omega_a}p_{\nu_b}q(\omega_a,\nu_b)$ is the probability of having a click event. The average pairing time, which denotes the time consumed to form a successful effective pairing is estimated as $T_{\rm{mean}}=\frac{\sum_{i=0}^{N_{T_c}-1}(i+1)\left(1-q_{\rm{tot}}\right)^iq_{\rm{tot}}}{F\sum_{i=0}^{N_{T_c}-1}\left(1-q_{\rm{tot}}\right)^iq_{\rm{tot}}}=\frac{1-N_{T_c}q_{\rm{tot}}\left(1/qT_c-1\right)}{Fq_{\rm{tot}}}$.  The number of successful pairing event pairs $S_{\left[k_a^{tot},k_b^{tot}\right]}$ needed in parameter estimates (except the set $S_{\left[2\omega_a,2\omega_b\right]}$) can be calculated as follows:
  
        \begin{widetext}
        \begin{equation}
            n_{\left[k_a^{\rm{tot}},k_b^{\rm{tot}}\right]}=n_{\rm{tot}}\sum_{k_a^e+k_a^l}\sum_{k_b^e+k_b^l}\left(\frac{p_{k_a}^ep_{k_b}^eq\left(k_a^e,k_b^e\right)}{q_{\rm{tot}}}\frac{p_{k_a}^lp_{k_b}^lq\left(k_a^l,k_b^l\right)}{q_{\rm{tot}}}\right).
        \end{equation}
        And the set $S_{\left[2\omega_a,2\omega_b\right]}$ contains the successful pairing events that need to consider the phase technologies, which is calculated as follows:
        \begin{equation}
            n_{\left[2\omega_a,2\omega_b\right]}=\frac{n_{\rm{tot}}}{M\pi}\int\limits_{0}^{2\pi}\left(\frac{p_{\omega_a}p_{\omega_b}q^{\theta}\left(\omega_a,\omega_b\right)}{q_{\rm{tot}}}\frac{p_{\omega_a}p_{\omega_b}q^{\theta}\left(\omega_a,\omega_b\right)}{q_{\rm{tot}}}\right)d\theta.
        \end{equation}
        M represents the number of intervals in which the phase is divided, $q^{\theta}\left(\omega_a,\omega_b\right)=q^{\theta,L}_{\left(\omega_a,\omega_b\right)}+q^{\theta,R}_{\left(\omega_a,\omega_b\right)}$.

        Although the phase is discrete modulated in the experiment, it is continuous and randomly distributed while reaching the measurement part because of the phase drift in the fiber. Therefor, the X-basis error number is calculating as
        
        \begin{equation}
            \begin{aligned}
                m_{\left[2\omega_a,2\omega_b\right]}&=\frac{n_{\rm{tot}}}{M\pi}\int\limits_{0}^{2\pi}\bigg\{\left(1-E_{\rm{HOM}}\right)\frac{p_{\omega_a}^2p_{\omega_b}^2\left[q_{\left(\omega_a,\omega_b\right)}^{\theta,L}q_{\left(\omega_a,\omega_b\right)}^{\theta+\delta,R}+q_{\left(\omega_a,\omega_b\right)}^{\theta,R}q_{\left(\omega_a,\omega_b\right)}^{\theta+\delta,L}\right]}{q_{\rm{tot}}^2}\\
                &+E_{\rm{HOM}}\frac{p_{\omega_a}^2p_{\omega_b}^2\left[q_{\left(\omega_a,\omega_b\right)}^{\theta,L}q_{\left(\omega_a,\omega_b\right)}^{\theta+\delta,L}+q_{\left(\omega_a,\omega_b\right)}^{\theta,R}q_{\left(\omega_a,\omega_b\right)}^{\theta+\delta,R}\right]}{q_{\rm{tot}}^2}\bigg\}d\theta,
            \end{aligned}
        \end{equation}
        where $E_{\rm{HOM}}$ is the interference mislignment error rate and $\delta=T_{\rm{mean}}\left(2\pi\delta\nu+\omega_{\rm{fib}}\right)$ is the phase mislignment error rate caused by the fiber phase drift $\omega_{\rm{fib}}$ and the difference $\delta\nu$ between the two laser frequencies. However, the frequency difference $\delta\nu$ between the two lasers in the simulation is usually small enough to be ignored.
        \end{widetext}

    \section{Key Rate}
	\begin{enumerate}
    
    \item The AMDI key rate obtained in the key distribution phase is
    \begin{equation}
    \begin{aligned}
    R_{\rm{key}}&=\frac{F}{N}\bigg\{\underline{s}_0^z+\underline{s}_{11}^z\left[1-H_2\left(\overline{\phi}_{11}^z\right)\right]\\
    &-\lambda_{\rm{EC}}-\log_2\frac{2}{\epsilon_{\rm{cor}}} -2\log_2\frac{2}{\epsilon'\hat{\epsilon}}-2\log_2\frac{1}{2\epsilon_{\rm{PA}}} \bigg\}.
    \end{aligned}
    \end{equation}
    \item The data consumed in the error correction process is
        \begin{equation}
            \begin{aligned}\lambda_{\rm{EC}}=n_{\left[\xi\left(\mu_a\right),\xi\left(\mu_b\right)\right]}fH_2\left(\frac{m_{\left[\xi\left(\mu_a\right),\xi\left(\mu_b\right)\right]}}{n_{\left[\xi\left(\mu_a\right),\xi\left(\mu_b\right)\right]}}\right).
            \end{aligned}
        \end{equation}

    \section{Some optimized parameter values}
    We also list the optimized parameter values at typical distance points for the  perfectly modulated signal state CSS-AMDI in Table \ref{tab 01}. The time windows are set as $N=10^{12}$. The parameters $\mu_a$, $\nu_a$, $\omega_a$, $p_{\xi(\mu_a)}$, $p_{\nu_a}$, $p_{\omega_a}$, and $T_c$ are intensity in Z basis used for generate $\xi(\mu_a)$, decoy state intensities $\nu_a$, decoy state intensities $\omega_a$, probability of choosing signal state $\xi(\mu_a)$, probability of choosing decoy state $\nu_a$, probability of choosing decoy state $\omega_a$, and time interval $T_c$. The corresponding parameters for Bob are $\mu_b=\mu_a$, $\nu_b=\nu_a$, $\omega_b=\omega_a$, $p_{\xi(\mu_b)}=p_{\xi(\mu_a)}$, $p_{\nu_b}=p_{\nu_a}$, and $p_{\omega_b}=p_{\omega_a}$.

    \end{enumerate}




\begin{thebibliography}{42}%
\makeatletter
\providecommand \@ifxundefined [1]{%
 \@ifx{#1\undefined}
}%
\providecommand \@ifnum [1]{%
 \ifnum #1\expandafter \@firstoftwo
 \else \expandafter \@secondoftwo
 \fi
}%
\providecommand \@ifx [1]{%
 \ifx #1\expandafter \@firstoftwo
 \else \expandafter \@secondoftwo
 \fi
}%
\providecommand \natexlab [1]{#1}%
\providecommand \enquote  [1]{``#1''}%
\providecommand \bibnamefont  [1]{#1}%
\providecommand \bibfnamefont [1]{#1}%
\providecommand \citenamefont [1]{#1}%
\providecommand \href@noop [0]{\@secondoftwo}%
\providecommand \href [0]{\begingroup \@sanitize@url \@href}%
\providecommand \@href[1]{\@@startlink{#1}\@@href}%
\providecommand \@@href[1]{\endgroup#1\@@endlink}%
\providecommand \@sanitize@url [0]{\catcode `\\12\catcode `\$12\catcode
  `\&12\catcode `\#12\catcode `\^12\catcode `\_12\catcode `\%12\relax}%
\providecommand \@@startlink[1]{}%
\providecommand \@@endlink[0]{}%
\providecommand \url  [0]{\begingroup\@sanitize@url \@url }%
\providecommand \@url [1]{\endgroup\@href {#1}{\urlprefix }}%
\providecommand \urlprefix  [0]{URL }%
\providecommand \Eprint [0]{\href }%
\providecommand \doibase [0]{https://doi.org/}%
\providecommand \selectlanguage [0]{\@gobble}%
\providecommand \bibinfo  [0]{\@secondoftwo}%
\providecommand \bibfield  [0]{\@secondoftwo}%
\providecommand \translation [1]{[#1]}%
\providecommand \BibitemOpen [0]{}%
\providecommand \bibitemStop [0]{}%
\providecommand \bibitemNoStop [0]{.\EOS\space}%
\providecommand \EOS [0]{\spacefactor3000\relax}%
\providecommand \BibitemShut  [1]{\csname bibitem#1\endcsname}%
\let\auto@bib@innerbib\@empty
\bibitem [{\citenamefont {Lo}\ \emph {et~al.}(2012)\citenamefont {Lo},
  \citenamefont {Curty},\ and\ \citenamefont {Qi}}]{lo2012measurement}%
  \BibitemOpen
  \bibfield  {author} {\bibinfo {author} {\bibfnamefont {H.-K.}\ \bibnamefont
  {Lo}}, \bibinfo {author} {\bibfnamefont {M.}~\bibnamefont {Curty}},\ and\
  \bibinfo {author} {\bibfnamefont {B.}~\bibnamefont {Qi}},\ }\bibfield
  {title} {\bibinfo {title} {Measurement-device-independent quantum key
  distribution},\ }\href@noop {} {\bibfield  {journal} {\bibinfo  {journal}
  {Phys. Rev. Lett.}\ }\textbf {\bibinfo {volume} {108}},\ \bibinfo {pages}
  {130503} (\bibinfo {year} {2012})}\BibitemShut {NoStop}%
\bibitem [{\citenamefont {Wang}(2013)}]{wang2013three}%
  \BibitemOpen
  \bibfield  {author} {\bibinfo {author} {\bibfnamefont {X.-B.}\ \bibnamefont
  {Wang}},\ }\bibfield  {title} {\bibinfo {title} {Three-intensity decoy-state
  method for device-independent quantum key distribution with basis-dependent
  errors},\ }\href@noop {} {\bibfield  {journal} {\bibinfo  {journal} {Phys.
  Rev. A}\ }\textbf {\bibinfo {volume} {87}},\ \bibinfo {pages} {012320}
  (\bibinfo {year} {2013})}\BibitemShut {NoStop}%
\bibitem [{\citenamefont {Yin}\ \emph {et~al.}(2016)\citenamefont {Yin},
  \citenamefont {Chen}, \citenamefont {Yu}, \citenamefont {Liu}, \citenamefont
  {You}, \citenamefont {Zhou}, \citenamefont {Chen}, \citenamefont {Mao},
  \citenamefont {Huang}, \citenamefont {Zhang}, \citenamefont {Chen},
  \citenamefont {Li}, \citenamefont {Nolan}, \citenamefont {Zhou},
  \citenamefont {Jiang}, \citenamefont {Wang}, \citenamefont {Zhang},
  \citenamefont {Wang},\ and\ \citenamefont {Pan}}]{yin2016measurement}%
  \BibitemOpen
  \bibfield  {author} {\bibinfo {author} {\bibfnamefont {H.-L.}\ \bibnamefont
  {Yin}}, \bibinfo {author} {\bibfnamefont {T.-Y.}\ \bibnamefont {Chen}},
  \bibinfo {author} {\bibfnamefont {Z.-W.}\ \bibnamefont {Yu}}, \bibinfo
  {author} {\bibfnamefont {H.}~\bibnamefont {Liu}}, \bibinfo {author}
  {\bibfnamefont {L.-X.}\ \bibnamefont {You}}, \bibinfo {author} {\bibfnamefont
  {Y.-H.}\ \bibnamefont {Zhou}}, \bibinfo {author} {\bibfnamefont {S.-J.}\
  \bibnamefont {Chen}}, \bibinfo {author} {\bibfnamefont {Y.}~\bibnamefont
  {Mao}}, \bibinfo {author} {\bibfnamefont {M.-Q.}\ \bibnamefont {Huang}},
  \bibinfo {author} {\bibfnamefont {W.-J.}\ \bibnamefont {Zhang}}, \bibinfo
  {author} {\bibfnamefont {H.}~\bibnamefont {Chen}}, \bibinfo {author}
  {\bibfnamefont {M.~J.}\ \bibnamefont {Li}}, \bibinfo {author} {\bibfnamefont
  {D.}~\bibnamefont {Nolan}}, \bibinfo {author} {\bibfnamefont
  {F.}~\bibnamefont {Zhou}}, \bibinfo {author} {\bibfnamefont {X.}~\bibnamefont
  {Jiang}}, \bibinfo {author} {\bibfnamefont {Z.}~\bibnamefont {Wang}},
  \bibinfo {author} {\bibfnamefont {Q.}~\bibnamefont {Zhang}}, \bibinfo
  {author} {\bibfnamefont {X.-B.}\ \bibnamefont {Wang}},\ and\ \bibinfo
  {author} {\bibfnamefont {J.-W.}\ \bibnamefont {Pan}},\ }\bibfield  {title}
  {\bibinfo {title} {Measurement-device-independent quantum key distribution
  over a 404 km optical fiber},\ }\href@noop {} {\bibfield  {journal} {\bibinfo
   {journal} {Phys. Rev. Lett.}\ }\textbf {\bibinfo {volume} {117}},\ \bibinfo
  {pages} {190501} (\bibinfo {year} {2016})}\BibitemShut {NoStop}%
\bibitem [{\citenamefont {Zhou}\ \emph {et~al.}(2016)\citenamefont {Zhou},
  \citenamefont {Yu},\ and\ \citenamefont {Wang}}]{zhou2016making}%
  \BibitemOpen
  \bibfield  {author} {\bibinfo {author} {\bibfnamefont {Y.-H.}\ \bibnamefont
  {Zhou}}, \bibinfo {author} {\bibfnamefont {Z.-W.}\ \bibnamefont {Yu}},\ and\
  \bibinfo {author} {\bibfnamefont {X.-B.}\ \bibnamefont {Wang}},\ }\bibfield
  {title} {\bibinfo {title} {Making the decoy-state
  measurement-device-independent quantum key distribution practically useful},\
  }\href@noop {} {\bibfield  {journal} {\bibinfo  {journal} {Phys. Rev. A}\
  }\textbf {\bibinfo {volume} {93}},\ \bibinfo {pages} {042324} (\bibinfo
  {year} {2016})}\BibitemShut {NoStop}%
\bibitem [{\citenamefont {Boaron}\ \emph {et~al.}(2018)\citenamefont {Boaron},
  \citenamefont {Boso}, \citenamefont {Rusca}, \citenamefont {Vulliez},
  \citenamefont {Autebert}, \citenamefont {Caloz}, \citenamefont {Perrenoud},
  \citenamefont {Gras}, \citenamefont {Bussi{\`e}res}, \citenamefont {Li} \emph
  {et~al.}}]{boaron2018secure}%
  \BibitemOpen
  \bibfield  {author} {\bibinfo {author} {\bibfnamefont {A.}~\bibnamefont
  {Boaron}}, \bibinfo {author} {\bibfnamefont {G.}~\bibnamefont {Boso}},
  \bibinfo {author} {\bibfnamefont {D.}~\bibnamefont {Rusca}}, \bibinfo
  {author} {\bibfnamefont {C.}~\bibnamefont {Vulliez}}, \bibinfo {author}
  {\bibfnamefont {C.}~\bibnamefont {Autebert}}, \bibinfo {author}
  {\bibfnamefont {M.}~\bibnamefont {Caloz}}, \bibinfo {author} {\bibfnamefont
  {M.}~\bibnamefont {Perrenoud}}, \bibinfo {author} {\bibfnamefont
  {G.}~\bibnamefont {Gras}}, \bibinfo {author} {\bibfnamefont {F.}~\bibnamefont
  {Bussi{\`e}res}}, \bibinfo {author} {\bibfnamefont {M.-J.}\ \bibnamefont
  {Li}}, \emph {et~al.},\ }\bibfield  {title} {\bibinfo {title} {Secure quantum
  key distribution over 421 km of optical fiber},\ }\href@noop {} {\bibfield
  {journal} {\bibinfo  {journal} {Phys. Rev. Lett.}\ }\textbf {\bibinfo
  {volume} {121}},\ \bibinfo {pages} {190502} (\bibinfo {year}
  {2018})}\BibitemShut {NoStop}%
\bibitem [{\citenamefont {Liu}\ \emph {et~al.}(2019)\citenamefont {Liu},
  \citenamefont {Yu}, \citenamefont {Zhang}, \citenamefont {Guan},
  \citenamefont {Chen}, \citenamefont {Zhang}, \citenamefont {Hu},
  \citenamefont {Li}, \citenamefont {Jiang}, \citenamefont {Lin}, \citenamefont
  {Chen}, \citenamefont {You}, \citenamefont {Wang}, \citenamefont {Wang},
  \citenamefont {Zhang},\ and\ \citenamefont {Pan}}]{liu2019experimental}%
  \BibitemOpen
  \bibfield  {author} {\bibinfo {author} {\bibfnamefont {Y.}~\bibnamefont
  {Liu}}, \bibinfo {author} {\bibfnamefont {Z.-W.}\ \bibnamefont {Yu}},
  \bibinfo {author} {\bibfnamefont {W.}~\bibnamefont {Zhang}}, \bibinfo
  {author} {\bibfnamefont {J.-Y.}\ \bibnamefont {Guan}}, \bibinfo {author}
  {\bibfnamefont {J.-P.}\ \bibnamefont {Chen}}, \bibinfo {author}
  {\bibfnamefont {C.}~\bibnamefont {Zhang}}, \bibinfo {author} {\bibfnamefont
  {X.-L.}\ \bibnamefont {Hu}}, \bibinfo {author} {\bibfnamefont
  {H.}~\bibnamefont {Li}}, \bibinfo {author} {\bibfnamefont {C.}~\bibnamefont
  {Jiang}}, \bibinfo {author} {\bibfnamefont {J.}~\bibnamefont {Lin}}, \bibinfo
  {author} {\bibfnamefont {T.-Y.}\ \bibnamefont {Chen}}, \bibinfo {author}
  {\bibfnamefont {L.}~\bibnamefont {You}}, \bibinfo {author} {\bibfnamefont
  {Z.}~\bibnamefont {Wang}}, \bibinfo {author} {\bibfnamefont {X.-B.}\
  \bibnamefont {Wang}}, \bibinfo {author} {\bibfnamefont {Q.}~\bibnamefont
  {Zhang}},\ and\ \bibinfo {author} {\bibfnamefont {J.-W.}\ \bibnamefont
  {Pan}},\ }\bibfield  {title} {\bibinfo {title} {Experimental twin-field
  quantum key distribution through sending or not sending},\ }\href@noop {}
  {\bibfield  {journal} {\bibinfo  {journal} {Phys. Rev. Lett.}\ }\textbf
  {\bibinfo {volume} {123}},\ \bibinfo {pages} {100505} (\bibinfo {year}
  {2019})}\BibitemShut {NoStop}%
\bibitem [{\citenamefont {Wei}\ \emph {et~al.}(2020)\citenamefont {Wei},
  \citenamefont {Li}, \citenamefont {Tan}, \citenamefont {Li}, \citenamefont
  {Min}, \citenamefont {Zhang}, \citenamefont {Li}, \citenamefont {You},
  \citenamefont {Wang}, \citenamefont {Jiang} \emph {et~al.}}]{wei2020high}%
  \BibitemOpen
  \bibfield  {author} {\bibinfo {author} {\bibfnamefont {K.}~\bibnamefont
  {Wei}}, \bibinfo {author} {\bibfnamefont {W.}~\bibnamefont {Li}}, \bibinfo
  {author} {\bibfnamefont {H.}~\bibnamefont {Tan}}, \bibinfo {author}
  {\bibfnamefont {Y.}~\bibnamefont {Li}}, \bibinfo {author} {\bibfnamefont
  {H.}~\bibnamefont {Min}}, \bibinfo {author} {\bibfnamefont {W.-J.}\
  \bibnamefont {Zhang}}, \bibinfo {author} {\bibfnamefont {H.}~\bibnamefont
  {Li}}, \bibinfo {author} {\bibfnamefont {L.}~\bibnamefont {You}}, \bibinfo
  {author} {\bibfnamefont {Z.}~\bibnamefont {Wang}}, \bibinfo {author}
  {\bibfnamefont {X.}~\bibnamefont {Jiang}}, \emph {et~al.},\ }\bibfield
  {title} {\bibinfo {title} {High-speed measurement-device-independent quantum
  key distribution with integrated silicon photonics},\ }\href@noop {}
  {\bibfield  {journal} {\bibinfo  {journal} {Phys. Rev. X}\ }\textbf {\bibinfo
  {volume} {10}},\ \bibinfo {pages} {031030} (\bibinfo {year}
  {2020})}\BibitemShut {NoStop}%
\bibitem [{\citenamefont {Liu}\ \emph {et~al.}(2021)\citenamefont {Liu},
  \citenamefont {Li}, \citenamefont {Xie}, \citenamefont {Weng}, \citenamefont
  {Gu}, \citenamefont {Cao}, \citenamefont {Lu}, \citenamefont {Li},
  \citenamefont {Yin},\ and\ \citenamefont {Chen}}]{liu2021homodyne}%
  \BibitemOpen
  \bibfield  {author} {\bibinfo {author} {\bibfnamefont {W.-B.}\ \bibnamefont
  {Liu}}, \bibinfo {author} {\bibfnamefont {C.-L.}\ \bibnamefont {Li}},
  \bibinfo {author} {\bibfnamefont {Y.-M.}\ \bibnamefont {Xie}}, \bibinfo
  {author} {\bibfnamefont {C.-X.}\ \bibnamefont {Weng}}, \bibinfo {author}
  {\bibfnamefont {J.}~\bibnamefont {Gu}}, \bibinfo {author} {\bibfnamefont
  {X.-Y.}\ \bibnamefont {Cao}}, \bibinfo {author} {\bibfnamefont {Y.-S.}\
  \bibnamefont {Lu}}, \bibinfo {author} {\bibfnamefont {B.-H.}\ \bibnamefont
  {Li}}, \bibinfo {author} {\bibfnamefont {H.-L.}\ \bibnamefont {Yin}},\ and\
  \bibinfo {author} {\bibfnamefont {Z.-B.}\ \bibnamefont {Chen}},\ }\bibfield
  {title} {\bibinfo {title} {Homodyne detection quadrature phase shift keying
  continuous-variable quantum key distribution with high excess noise
  tolerance},\ }\href@noop {} {\bibfield  {journal} {\bibinfo  {journal} {PRX
  Quantum}\ }\textbf {\bibinfo {volume} {2}},\ \bibinfo {pages} {040334}
  (\bibinfo {year} {2021})}\BibitemShut {NoStop}%
\bibitem [{\citenamefont {Wang}\ \emph {et~al.}(2022)\citenamefont {Wang},
  \citenamefont {Yin}, \citenamefont {He}, \citenamefont {Chen}, \citenamefont
  {Wang}, \citenamefont {Ye}, \citenamefont {Zhou}, \citenamefont {Fan-Yuan},
  \citenamefont {Wang}, \citenamefont {Chen}, \citenamefont {Zhu},
  \citenamefont {V.~Morozov}, \citenamefont {V.~Divochiy}, \citenamefont
  {Zhou}, \citenamefont {Guo},\ and\ \citenamefont {Han}}]{wang2022twin}%
  \BibitemOpen
  \bibfield  {author} {\bibinfo {author} {\bibfnamefont {S.}~\bibnamefont
  {Wang}}, \bibinfo {author} {\bibfnamefont {Z.-Q.}\ \bibnamefont {Yin}},
  \bibinfo {author} {\bibfnamefont {D.-Y.}\ \bibnamefont {He}}, \bibinfo
  {author} {\bibfnamefont {W.}~\bibnamefont {Chen}}, \bibinfo {author}
  {\bibfnamefont {R.-Q.}\ \bibnamefont {Wang}}, \bibinfo {author}
  {\bibfnamefont {P.}~\bibnamefont {Ye}}, \bibinfo {author} {\bibfnamefont
  {Y.}~\bibnamefont {Zhou}}, \bibinfo {author} {\bibfnamefont {G.-J.}\
  \bibnamefont {Fan-Yuan}}, \bibinfo {author} {\bibfnamefont {F.-X.}\
  \bibnamefont {Wang}}, \bibinfo {author} {\bibfnamefont {W.}~\bibnamefont
  {Chen}}, \bibinfo {author} {\bibfnamefont {Y.-G.}\ \bibnamefont {Zhu}},
  \bibinfo {author} {\bibfnamefont {P.}~\bibnamefont {V.~Morozov}}, \bibinfo
  {author} {\bibfnamefont {A.}~\bibnamefont {V.~Divochiy}}, \bibinfo {author}
  {\bibfnamefont {Z.}~\bibnamefont {Zhou}}, \bibinfo {author} {\bibfnamefont
  {G.-C.}\ \bibnamefont {Guo}},\ and\ \bibinfo {author} {\bibfnamefont {Z.-F.}\
  \bibnamefont {Han}},\ }\bibfield  {title} {\bibinfo {title} {Twin-field
  quantum key distribution over 830-km fibre},\ }\href@noop {} {\bibfield
  {journal} {\bibinfo  {journal} {Nat. Photon.}\ }\textbf {\bibinfo {volume}
  {16}},\ \bibinfo {pages} {154} (\bibinfo {year} {2022})}\BibitemShut
  {NoStop}%
\bibitem [{\citenamefont {Gu}\ \emph {et~al.}(2022)\citenamefont {Gu},
  \citenamefont {Cao}, \citenamefont {Fu}, \citenamefont {He}, \citenamefont
  {Yin}, \citenamefont {Yin},\ and\ \citenamefont {Chen}}]{gu2022experimental}%
  \BibitemOpen
  \bibfield  {author} {\bibinfo {author} {\bibfnamefont {J.}~\bibnamefont
  {Gu}}, \bibinfo {author} {\bibfnamefont {X.-Y.}\ \bibnamefont {Cao}},
  \bibinfo {author} {\bibfnamefont {Y.}~\bibnamefont {Fu}}, \bibinfo {author}
  {\bibfnamefont {Z.-W.}\ \bibnamefont {He}}, \bibinfo {author} {\bibfnamefont
  {Z.-J.}\ \bibnamefont {Yin}}, \bibinfo {author} {\bibfnamefont {H.-L.}\
  \bibnamefont {Yin}},\ and\ \bibinfo {author} {\bibfnamefont {Z.-B.}\
  \bibnamefont {Chen}},\ }\bibfield  {title} {\bibinfo {title} {Experimental
  measurement-device-independent type quantum key distribution with flawed and
  correlated sources},\ }\href@noop {} {\bibfield  {journal} {\bibinfo
  {journal} {Science Bulletin}\ }\textbf {\bibinfo {volume} {67}},\ \bibinfo
  {pages} {2167} (\bibinfo {year} {2022})}\BibitemShut {NoStop}%
\bibitem [{\citenamefont {Zhou}\ \emph
  {et~al.}(2023{\natexlab{a}})\citenamefont {Zhou}, \citenamefont {Lin},
  \citenamefont {Jing},\ and\ \citenamefont {Yuan}}]{zhou2023twin}%
  \BibitemOpen
  \bibfield  {author} {\bibinfo {author} {\bibfnamefont {L.}~\bibnamefont
  {Zhou}}, \bibinfo {author} {\bibfnamefont {J.}~\bibnamefont {Lin}}, \bibinfo
  {author} {\bibfnamefont {Y.}~\bibnamefont {Jing}},\ and\ \bibinfo {author}
  {\bibfnamefont {Z.}~\bibnamefont {Yuan}},\ }\bibfield  {title} {\bibinfo
  {title} {Twin-field quantum key distribution without optical frequency
  dissemination},\ }\href@noop {} {\bibfield  {journal} {\bibinfo  {journal}
  {Nature Communications}\ }\textbf {\bibinfo {volume} {14}},\ \bibinfo {pages}
  {928} (\bibinfo {year} {2023}{\natexlab{a}})}\BibitemShut {NoStop}%
\bibitem [{\citenamefont {Lydersen}\ \emph {et~al.}(2010)\citenamefont
  {Lydersen}, \citenamefont {Wiechers}, \citenamefont {Wittmann}, \citenamefont
  {Elser}, \citenamefont {Skaar},\ and\ \citenamefont
  {Makarov}}]{lydersen2010hacking}%
  \BibitemOpen
  \bibfield  {author} {\bibinfo {author} {\bibfnamefont {L.}~\bibnamefont
  {Lydersen}}, \bibinfo {author} {\bibfnamefont {C.}~\bibnamefont {Wiechers}},
  \bibinfo {author} {\bibfnamefont {C.}~\bibnamefont {Wittmann}}, \bibinfo
  {author} {\bibfnamefont {D.}~\bibnamefont {Elser}}, \bibinfo {author}
  {\bibfnamefont {J.}~\bibnamefont {Skaar}},\ and\ \bibinfo {author}
  {\bibfnamefont {V.}~\bibnamefont {Makarov}},\ }\bibfield  {title} {\bibinfo
  {title} {Hacking commercial quantum cryptography systems by tailored bright
  illumination},\ }\href@noop {} {\bibfield  {journal} {\bibinfo  {journal}
  {Nat. Photonics}\ }\textbf {\bibinfo {volume} {4}},\ \bibinfo {pages} {686}
  (\bibinfo {year} {2010})}\BibitemShut {NoStop}%
\bibitem [{\citenamefont {Tang}\ \emph {et~al.}(2013)\citenamefont {Tang},
  \citenamefont {Yin}, \citenamefont {Ma}, \citenamefont {Fung}, \citenamefont
  {Liu}, \citenamefont {Yong}, \citenamefont {Chen}, \citenamefont {Peng},
  \citenamefont {Chen},\ and\ \citenamefont {Pan}}]{tang2013source}%
  \BibitemOpen
  \bibfield  {author} {\bibinfo {author} {\bibfnamefont {Y.-L.}\ \bibnamefont
  {Tang}}, \bibinfo {author} {\bibfnamefont {H.-L.}\ \bibnamefont {Yin}},
  \bibinfo {author} {\bibfnamefont {X.}~\bibnamefont {Ma}}, \bibinfo {author}
  {\bibfnamefont {C.-H.~F.}\ \bibnamefont {Fung}}, \bibinfo {author}
  {\bibfnamefont {Y.}~\bibnamefont {Liu}}, \bibinfo {author} {\bibfnamefont
  {H.-L.}\ \bibnamefont {Yong}}, \bibinfo {author} {\bibfnamefont {T.-Y.}\
  \bibnamefont {Chen}}, \bibinfo {author} {\bibfnamefont {C.-Z.}\ \bibnamefont
  {Peng}}, \bibinfo {author} {\bibfnamefont {Z.-B.}\ \bibnamefont {Chen}},\
  and\ \bibinfo {author} {\bibfnamefont {J.-W.}\ \bibnamefont {Pan}},\
  }\bibfield  {title} {\bibinfo {title} {Source attack of decoy-state quantum
  key distribution using phase information},\ }\href@noop {} {\bibfield
  {journal} {\bibinfo  {journal} {Physical Review A}\ }\textbf {\bibinfo
  {volume} {88}},\ \bibinfo {pages} {022308} (\bibinfo {year}
  {2013})}\BibitemShut {NoStop}%
\bibitem [{\citenamefont {Xu}\ \emph {et~al.}(2020{\natexlab{a}})\citenamefont
  {Xu}, \citenamefont {Ma}, \citenamefont {Zhang}, \citenamefont {Lo},\ and\
  \citenamefont {Pan}}]{xu2020secure}%
  \BibitemOpen
  \bibfield  {author} {\bibinfo {author} {\bibfnamefont {F.}~\bibnamefont
  {Xu}}, \bibinfo {author} {\bibfnamefont {X.}~\bibnamefont {Ma}}, \bibinfo
  {author} {\bibfnamefont {Q.}~\bibnamefont {Zhang}}, \bibinfo {author}
  {\bibfnamefont {H.-K.}\ \bibnamefont {Lo}},\ and\ \bibinfo {author}
  {\bibfnamefont {J.-W.}\ \bibnamefont {Pan}},\ }\bibfield  {title} {\bibinfo
  {title} {Secure quantum key distribution with realistic devices},\
  }\href@noop {} {\bibfield  {journal} {\bibinfo  {journal} {Rev. Mod. Phys.}\
  }\textbf {\bibinfo {volume} {92}},\ \bibinfo {pages} {025002} (\bibinfo
  {year} {2020}{\natexlab{a}})}\BibitemShut {NoStop}%
\bibitem [{\citenamefont {Pirandola}\ \emph {et~al.}(2020)\citenamefont
  {Pirandola}, \citenamefont {Andersen}, \citenamefont {Banchi}, \citenamefont
  {Berta}, \citenamefont {Bunandar}, \citenamefont {Colbeck}, \citenamefont
  {Englund}, \citenamefont {Gehring}, \citenamefont {Lupo}, \citenamefont
  {Ottaviani} \emph {et~al.}}]{pirandola2020advances}%
  \BibitemOpen
  \bibfield  {author} {\bibinfo {author} {\bibfnamefont {S.}~\bibnamefont
  {Pirandola}}, \bibinfo {author} {\bibfnamefont {U.~L.}\ \bibnamefont
  {Andersen}}, \bibinfo {author} {\bibfnamefont {L.}~\bibnamefont {Banchi}},
  \bibinfo {author} {\bibfnamefont {M.}~\bibnamefont {Berta}}, \bibinfo
  {author} {\bibfnamefont {D.}~\bibnamefont {Bunandar}}, \bibinfo {author}
  {\bibfnamefont {R.}~\bibnamefont {Colbeck}}, \bibinfo {author} {\bibfnamefont
  {D.}~\bibnamefont {Englund}}, \bibinfo {author} {\bibfnamefont
  {T.}~\bibnamefont {Gehring}}, \bibinfo {author} {\bibfnamefont
  {C.}~\bibnamefont {Lupo}}, \bibinfo {author} {\bibfnamefont {C.}~\bibnamefont
  {Ottaviani}}, \emph {et~al.},\ }\bibfield  {title} {\bibinfo {title}
  {Advances in quantum cryptography},\ }\href@noop {} {\bibfield  {journal}
  {\bibinfo  {journal} {Adv. Opt. Photonics}\ }\textbf {\bibinfo {volume}
  {12}},\ \bibinfo {pages} {1012} (\bibinfo {year} {2020})}\BibitemShut
  {NoStop}%
\bibitem [{\citenamefont {Braunstein}\ and\ \citenamefont
  {Pirandola}(2012)}]{braunstein2012side}%
  \BibitemOpen
  \bibfield  {author} {\bibinfo {author} {\bibfnamefont {S.~L.}\ \bibnamefont
  {Braunstein}}\ and\ \bibinfo {author} {\bibfnamefont {S.}~\bibnamefont
  {Pirandola}},\ }\bibfield  {title} {\bibinfo {title} {Side-channel-free
  quantum key distribution},\ }\href@noop {} {\bibfield  {journal} {\bibinfo
  {journal} {Phys. Rev. Lett.}\ }\textbf {\bibinfo {volume} {108}},\ \bibinfo
  {pages} {130502} (\bibinfo {year} {2012})}\BibitemShut {NoStop}%
\bibitem [{\citenamefont {Lucamarini}\ \emph {et~al.}(2018)\citenamefont
  {Lucamarini}, \citenamefont {Yuan}, \citenamefont {Dynes},\ and\
  \citenamefont {Shields}}]{lucamarini2018overcoming}%
  \BibitemOpen
  \bibfield  {author} {\bibinfo {author} {\bibfnamefont {M.}~\bibnamefont
  {Lucamarini}}, \bibinfo {author} {\bibfnamefont {Z.~L.}\ \bibnamefont
  {Yuan}}, \bibinfo {author} {\bibfnamefont {J.~F.}\ \bibnamefont {Dynes}},\
  and\ \bibinfo {author} {\bibfnamefont {A.~J.}\ \bibnamefont {Shields}},\
  }\bibfield  {title} {\bibinfo {title} {Overcoming the rate--distance limit of
  quantum key distribution without quantum repeaters},\ }\href@noop {}
  {\bibfield  {journal} {\bibinfo  {journal} {Nature}\ }\textbf {\bibinfo
  {volume} {557}},\ \bibinfo {pages} {400} (\bibinfo {year}
  {2018})}\BibitemShut {NoStop}%
\bibitem [{\citenamefont {Xu}\ \emph {et~al.}(2020{\natexlab{b}})\citenamefont
  {Xu}, \citenamefont {Yu}, \citenamefont {Jiang}, \citenamefont {Hu},\ and\
  \citenamefont {Wang}}]{xu2020sending}%
  \BibitemOpen
  \bibfield  {author} {\bibinfo {author} {\bibfnamefont {H.}~\bibnamefont
  {Xu}}, \bibinfo {author} {\bibfnamefont {Z.-W.}\ \bibnamefont {Yu}}, \bibinfo
  {author} {\bibfnamefont {C.}~\bibnamefont {Jiang}}, \bibinfo {author}
  {\bibfnamefont {X.-L.}\ \bibnamefont {Hu}},\ and\ \bibinfo {author}
  {\bibfnamefont {X.-B.}\ \bibnamefont {Wang}},\ }\bibfield  {title} {\bibinfo
  {title} {Sending-or-not-sending twin-field quantum key distribution: Breaking
  the direct transmission key rate},\ }\href@noop {} {\bibfield  {journal}
  {\bibinfo  {journal} {Phys. Rev. A}\ }\textbf {\bibinfo {volume} {101}},\
  \bibinfo {pages} {042330} (\bibinfo {year} {2020}{\natexlab{b}})}\BibitemShut
  {NoStop}%
\bibitem [{\citenamefont {Xie}\ \emph {et~al.}(2022)\citenamefont {Xie},
  \citenamefont {Lu}, \citenamefont {Weng}, \citenamefont {Cao}, \citenamefont
  {Jia}, \citenamefont {Bao}, \citenamefont {Wang}, \citenamefont {Fu},
  \citenamefont {Yin},\ and\ \citenamefont {Chen}}]{xie2022breaking}%
  \BibitemOpen
  \bibfield  {author} {\bibinfo {author} {\bibfnamefont {Y.-M.}\ \bibnamefont
  {Xie}}, \bibinfo {author} {\bibfnamefont {Y.-S.}\ \bibnamefont {Lu}},
  \bibinfo {author} {\bibfnamefont {C.-X.}\ \bibnamefont {Weng}}, \bibinfo
  {author} {\bibfnamefont {X.-Y.}\ \bibnamefont {Cao}}, \bibinfo {author}
  {\bibfnamefont {Z.-Y.}\ \bibnamefont {Jia}}, \bibinfo {author} {\bibfnamefont
  {Y.}~\bibnamefont {Bao}}, \bibinfo {author} {\bibfnamefont {Y.}~\bibnamefont
  {Wang}}, \bibinfo {author} {\bibfnamefont {Y.}~\bibnamefont {Fu}}, \bibinfo
  {author} {\bibfnamefont {H.-L.}\ \bibnamefont {Yin}},\ and\ \bibinfo {author}
  {\bibfnamefont {Z.-B.}\ \bibnamefont {Chen}},\ }\bibfield  {title} {\bibinfo
  {title} {Breaking the rate-loss bound of quantum key distribution with
  asynchronous two-photon interference},\ }\href@noop {} {\bibfield  {journal}
  {\bibinfo  {journal} {PRX Quantum}\ }\textbf {\bibinfo {volume} {3}},\
  \bibinfo {pages} {020315} (\bibinfo {year} {2022})}\BibitemShut {NoStop}%
\bibitem [{\citenamefont {Zeng}\ \emph {et~al.}(2022)\citenamefont {Zeng},
  \citenamefont {Zhou}, \citenamefont {Wu},\ and\ \citenamefont
  {Ma}}]{zeng2022mode}%
  \BibitemOpen
  \bibfield  {author} {\bibinfo {author} {\bibfnamefont {P.}~\bibnamefont
  {Zeng}}, \bibinfo {author} {\bibfnamefont {H.}~\bibnamefont {Zhou}}, \bibinfo
  {author} {\bibfnamefont {W.}~\bibnamefont {Wu}},\ and\ \bibinfo {author}
  {\bibfnamefont {X.}~\bibnamefont {Ma}},\ }\bibfield  {title} {\bibinfo
  {title} {Mode-pairing quantum key distribution},\ }\href@noop {} {\bibfield
  {journal} {\bibinfo  {journal} {Nature Communications}\ }\textbf {\bibinfo
  {volume} {13}},\ \bibinfo {pages} {3903} (\bibinfo {year}
  {2022})}\BibitemShut {NoStop}%
\bibitem [{\citenamefont {Jiang}\ \emph {et~al.}(2023)\citenamefont {Jiang},
  \citenamefont {Yu}, \citenamefont {Hu},\ and\ \citenamefont
  {Wang}}]{10.1093/nsr/nwac186}%
  \BibitemOpen
  \bibfield  {author} {\bibinfo {author} {\bibfnamefont {C.}~\bibnamefont
  {Jiang}}, \bibinfo {author} {\bibfnamefont {Z.-W.}\ \bibnamefont {Yu}},
  \bibinfo {author} {\bibfnamefont {X.-L.}\ \bibnamefont {Hu}},\ and\ \bibinfo
  {author} {\bibfnamefont {X.-B.}\ \bibnamefont {Wang}},\ }\bibfield  {title}
  {\bibinfo {title} {{Robust twin-field quantum key distribution through
  sending or not sending}},\ }\href {https://doi.org/10.1093/nsr/nwac186}
  {\bibfield  {journal} {\bibinfo  {journal} {Natl. Sci. Rev.}\ }\textbf
  {\bibinfo {volume} {10}},\ \bibinfo {pages} {nwac186} (\bibinfo {year}
  {2023})},\ \Eprint
  {https://arxiv.org/abs/https://academic.oup.com/nsr/article-pdf/10/4/nwac186/50033122/nwac186.pdf}
  {https://academic.oup.com/nsr/article-pdf/10/4/nwac186/50033122/nwac186.pdf}
  \BibitemShut {NoStop}%
\bibitem [{\citenamefont {Xie}\ \emph {et~al.}(2023)\citenamefont {Xie},
  \citenamefont {Bai}, \citenamefont {Lu}, \citenamefont {Weng}, \citenamefont
  {Yin},\ and\ \citenamefont {Chen}}]{xie2023advantages}%
  \BibitemOpen
  \bibfield  {author} {\bibinfo {author} {\bibfnamefont {Y.-M.}\ \bibnamefont
  {Xie}}, \bibinfo {author} {\bibfnamefont {J.-L.}\ \bibnamefont {Bai}},
  \bibinfo {author} {\bibfnamefont {Y.-S.}\ \bibnamefont {Lu}}, \bibinfo
  {author} {\bibfnamefont {C.-X.}\ \bibnamefont {Weng}}, \bibinfo {author}
  {\bibfnamefont {H.-L.}\ \bibnamefont {Yin}},\ and\ \bibinfo {author}
  {\bibfnamefont {Z.-B.}\ \bibnamefont {Chen}},\ }\bibfield  {title} {\bibinfo
  {title} {Advantages of asynchronous measurement-device-independent quantum
  key distribution in intercity networks},\ }\href@noop {} {\bibfield
  {journal} {\bibinfo  {journal} {arXiv preprint arXiv:2302.14349}\ } (\bibinfo
  {year} {2023})}\BibitemShut {NoStop}%
\bibitem [{\citenamefont {Ma}\ \emph {et~al.}(2018)\citenamefont {Ma},
  \citenamefont {Zeng},\ and\ \citenamefont {Zhou}}]{ma2018phase}%
  \BibitemOpen
  \bibfield  {author} {\bibinfo {author} {\bibfnamefont {X.}~\bibnamefont
  {Ma}}, \bibinfo {author} {\bibfnamefont {P.}~\bibnamefont {Zeng}},\ and\
  \bibinfo {author} {\bibfnamefont {H.}~\bibnamefont {Zhou}},\ }\bibfield
  {title} {\bibinfo {title} {Phase-matching quantum key distribution},\
  }\href@noop {} {\bibfield  {journal} {\bibinfo  {journal} {Phys. Rev.X}\
  }\textbf {\bibinfo {volume} {8}},\ \bibinfo {pages} {031043} (\bibinfo {year}
  {2018})}\BibitemShut {NoStop}%
\bibitem [{\citenamefont {Wang}\ \emph {et~al.}(2018)\citenamefont {Wang},
  \citenamefont {Yu},\ and\ \citenamefont {Hu}}]{Wang2018twin}%
  \BibitemOpen
  \bibfield  {author} {\bibinfo {author} {\bibfnamefont {X.-B.}\ \bibnamefont
  {Wang}}, \bibinfo {author} {\bibfnamefont {Z.-W.}\ \bibnamefont {Yu}},\ and\
  \bibinfo {author} {\bibfnamefont {X.-L.}\ \bibnamefont {Hu}},\ }\bibfield
  {title} {\bibinfo {title} {Twin-field quantum key distribution with large
  misalignment error},\ }\href@noop {} {\bibfield  {journal} {\bibinfo
  {journal} {Phys. Rev. A}\ }\textbf {\bibinfo {volume} {98}},\ \bibinfo
  {pages} {062323} (\bibinfo {year} {2018})}\BibitemShut {NoStop}%
\bibitem [{\citenamefont {Cui}\ \emph {et~al.}(2019)\citenamefont {Cui},
  \citenamefont {Yin}, \citenamefont {Wang}, \citenamefont {Chen},
  \citenamefont {Wang}, \citenamefont {Guo},\ and\ \citenamefont
  {Han}}]{Cui2019Twin}%
  \BibitemOpen
  \bibfield  {author} {\bibinfo {author} {\bibfnamefont {C.}~\bibnamefont
  {Cui}}, \bibinfo {author} {\bibfnamefont {Z.-Q.}\ \bibnamefont {Yin}},
  \bibinfo {author} {\bibfnamefont {R.}~\bibnamefont {Wang}}, \bibinfo {author}
  {\bibfnamefont {W.}~\bibnamefont {Chen}}, \bibinfo {author} {\bibfnamefont
  {S.}~\bibnamefont {Wang}}, \bibinfo {author} {\bibfnamefont {G.-C.}\
  \bibnamefont {Guo}},\ and\ \bibinfo {author} {\bibfnamefont {Z.-F.}\
  \bibnamefont {Han}},\ }\bibfield  {title} {\bibinfo {title} {Twin-field
  quantum key distribution without phase postselection},\ }\href@noop {}
  {\bibfield  {journal} {\bibinfo  {journal} {Phys. Rev. Applied}\ }\textbf
  {\bibinfo {volume} {11}},\ \bibinfo {pages} {034053} (\bibinfo {year}
  {2019})}\BibitemShut {NoStop}%
\bibitem [{\citenamefont {Curty}\ \emph {et~al.}(2019)\citenamefont {Curty},
  \citenamefont {Azuma},\ and\ \citenamefont {Lo}}]{curty2019simple}%
  \BibitemOpen
  \bibfield  {author} {\bibinfo {author} {\bibfnamefont {M.}~\bibnamefont
  {Curty}}, \bibinfo {author} {\bibfnamefont {K.}~\bibnamefont {Azuma}},\ and\
  \bibinfo {author} {\bibfnamefont {H.-K.}\ \bibnamefont {Lo}},\ }\bibfield
  {title} {\bibinfo {title} {Simple security proof of twin-field type quantum
  key distribution protocol},\ }\href@noop {} {\bibfield  {journal} {\bibinfo
  {journal} {npj Quantum Inf.}\ }\textbf {\bibinfo {volume} {5}},\ \bibinfo
  {pages} {64} (\bibinfo {year} {2019})}\BibitemShut {NoStop}%
\bibitem [{\citenamefont {Maeda}\ \emph {et~al.}(2019)\citenamefont {Maeda},
  \citenamefont {Sasaki},\ and\ \citenamefont
  {Koashi}}]{maeda2019repeaterless}%
  \BibitemOpen
  \bibfield  {author} {\bibinfo {author} {\bibfnamefont {K.}~\bibnamefont
  {Maeda}}, \bibinfo {author} {\bibfnamefont {T.}~\bibnamefont {Sasaki}},\ and\
  \bibinfo {author} {\bibfnamefont {M.}~\bibnamefont {Koashi}},\ }\bibfield
  {title} {\bibinfo {title} {Repeaterless quantum key distribution with
  efficient finite-key analysis overcoming the rate-distance limit},\
  }\href@noop {} {\bibfield  {journal} {\bibinfo  {journal} {Nat. Commun.}\
  }\textbf {\bibinfo {volume} {10}},\ \bibinfo {pages} {3140} (\bibinfo {year}
  {2019})}\BibitemShut {NoStop}%
\bibitem [{\citenamefont {Yin}\ and\ \citenamefont
  {Chen}(2019)}]{yin2019finite}%
  \BibitemOpen
  \bibfield  {author} {\bibinfo {author} {\bibfnamefont {H.-L.}\ \bibnamefont
  {Yin}}\ and\ \bibinfo {author} {\bibfnamefont {Z.-B.}\ \bibnamefont {Chen}},\
  }\bibfield  {title} {\bibinfo {title} {Finite-key analysis for twin-field
  quantum key distribution with composable security},\ }\href@noop {}
  {\bibfield  {journal} {\bibinfo  {journal} {Sci. Rep.}\ }\textbf {\bibinfo
  {volume} {9}},\ \bibinfo {pages} {17113} (\bibinfo {year}
  {2019})}\BibitemShut {NoStop}%
\bibitem [{\citenamefont {Pirandola}\ \emph {et~al.}(2017)\citenamefont
  {Pirandola}, \citenamefont {Laurenza}, \citenamefont {Ottaviani},\ and\
  \citenamefont {Banchi}}]{pirandola2017fundamental}%
  \BibitemOpen
  \bibfield  {author} {\bibinfo {author} {\bibfnamefont {S.}~\bibnamefont
  {Pirandola}}, \bibinfo {author} {\bibfnamefont {R.}~\bibnamefont {Laurenza}},
  \bibinfo {author} {\bibfnamefont {C.}~\bibnamefont {Ottaviani}},\ and\
  \bibinfo {author} {\bibfnamefont {L.}~\bibnamefont {Banchi}},\ }\bibfield
  {title} {\bibinfo {title} {Fundamental limits of repeaterless quantum
  communications},\ }\href@noop {} {\bibfield  {journal} {\bibinfo  {journal}
  {Nat. Commun.}\ }\textbf {\bibinfo {volume} {8}},\ \bibinfo {pages} {15043}
  (\bibinfo {year} {2017})}\BibitemShut {NoStop}%
\bibitem [{\citenamefont {Zhu}\ \emph {et~al.}(2023)\citenamefont {Zhu},
  \citenamefont {Huang}, \citenamefont {Liu}, \citenamefont {Zeng},
  \citenamefont {Zou}, \citenamefont {Dai}, \citenamefont {Tang}, \citenamefont
  {Li}, \citenamefont {You}, \citenamefont {Wang}, \citenamefont {Chen},
  \citenamefont {Ma}, \citenamefont {Chen},\ and\ \citenamefont
  {Pan}}]{Zhu2023exp}%
  \BibitemOpen
  \bibfield  {author} {\bibinfo {author} {\bibfnamefont {H.-T.}\ \bibnamefont
  {Zhu}}, \bibinfo {author} {\bibfnamefont {Y.}~\bibnamefont {Huang}}, \bibinfo
  {author} {\bibfnamefont {H.}~\bibnamefont {Liu}}, \bibinfo {author}
  {\bibfnamefont {P.}~\bibnamefont {Zeng}}, \bibinfo {author} {\bibfnamefont
  {M.}~\bibnamefont {Zou}}, \bibinfo {author} {\bibfnamefont {Y.}~\bibnamefont
  {Dai}}, \bibinfo {author} {\bibfnamefont {S.}~\bibnamefont {Tang}}, \bibinfo
  {author} {\bibfnamefont {H.}~\bibnamefont {Li}}, \bibinfo {author}
  {\bibfnamefont {L.}~\bibnamefont {You}}, \bibinfo {author} {\bibfnamefont
  {Z.}~\bibnamefont {Wang}}, \bibinfo {author} {\bibfnamefont {Y.-A.}\
  \bibnamefont {Chen}}, \bibinfo {author} {\bibfnamefont {X.}~\bibnamefont
  {Ma}}, \bibinfo {author} {\bibfnamefont {T.-Y.}\ \bibnamefont {Chen}},\ and\
  \bibinfo {author} {\bibfnamefont {J.-W.}\ \bibnamefont {Pan}},\ }\bibfield
  {title} {\bibinfo {title} {Experimental mode-pairing
  measurement-device-independent quantum key distribution without global phase
  locking},\ }\href@noop {} {\bibfield  {journal} {\bibinfo  {journal} {Phys.
  Rev. Lett.}\ }\textbf {\bibinfo {volume} {130}},\ \bibinfo {pages} {030801}
  (\bibinfo {year} {2023})}\BibitemShut {NoStop}%
\bibitem [{\citenamefont {Zhou}\ \emph
  {et~al.}(2023{\natexlab{b}})\citenamefont {Zhou}, \citenamefont {Lin},
  \citenamefont {Xie}, \citenamefont {Lu}, \citenamefont {Jing}, \citenamefont
  {Yin},\ and\ \citenamefont {Yuan}}]{zhou2022experimental}%
  \BibitemOpen
  \bibfield  {author} {\bibinfo {author} {\bibfnamefont {L.}~\bibnamefont
  {Zhou}}, \bibinfo {author} {\bibfnamefont {J.}~\bibnamefont {Lin}}, \bibinfo
  {author} {\bibfnamefont {Y.-M.}\ \bibnamefont {Xie}}, \bibinfo {author}
  {\bibfnamefont {Y.-S.}\ \bibnamefont {Lu}}, \bibinfo {author} {\bibfnamefont
  {Y.}~\bibnamefont {Jing}}, \bibinfo {author} {\bibfnamefont {H.-L.}\
  \bibnamefont {Yin}},\ and\ \bibinfo {author} {\bibfnamefont {Z.}~\bibnamefont
  {Yuan}},\ }\bibfield  {title} {\bibinfo {title} {Experimental quantum
  communication overcomes the rate-loss limit without global phase tracking},\
  }\href@noop {} {\bibfield  {journal} {\bibinfo  {journal} {arXiv preprint
  arXiv:2212.14190, accepted by Phys. Rev. Lett.}\ } (\bibinfo {year}
  {2023}{\natexlab{b}})}\BibitemShut {NoStop}%
\bibitem [{\citenamefont {Yin}\ \emph {et~al.}(2014)\citenamefont {Yin},
  \citenamefont {Cao}, \citenamefont {Fu}, \citenamefont {Tang}, \citenamefont
  {Liu}, \citenamefont {Chen},\ and\ \citenamefont {Chen}}]{yin2014long}%
  \BibitemOpen
  \bibfield  {author} {\bibinfo {author} {\bibfnamefont {H.-L.}\ \bibnamefont
  {Yin}}, \bibinfo {author} {\bibfnamefont {W.-F.}\ \bibnamefont {Cao}},
  \bibinfo {author} {\bibfnamefont {Y.}~\bibnamefont {Fu}}, \bibinfo {author}
  {\bibfnamefont {Y.-L.}\ \bibnamefont {Tang}}, \bibinfo {author}
  {\bibfnamefont {Y.}~\bibnamefont {Liu}}, \bibinfo {author} {\bibfnamefont
  {T.-Y.}\ \bibnamefont {Chen}},\ and\ \bibinfo {author} {\bibfnamefont
  {Z.-B.}\ \bibnamefont {Chen}},\ }\bibfield  {title} {\bibinfo {title}
  {Long-distance measurement-device-independent quantum key distribution with
  coherent-state superpositions},\ }\href@noop {} {\bibfield  {journal}
  {\bibinfo  {journal} {Opt. Lett.}\ }\textbf {\bibinfo {volume} {39}},\
  \bibinfo {pages} {5451} (\bibinfo {year} {2014})}\BibitemShut {NoStop}%
\bibitem [{\citenamefont {Zhang}\ \emph {et~al.}(2019)\citenamefont {Zhang},
  \citenamefont {Zhang},\ and\ \citenamefont {Wang}}]{zhang2019twin}%
  \BibitemOpen
  \bibfield  {author} {\bibinfo {author} {\bibfnamefont {C.-H.}\ \bibnamefont
  {Zhang}}, \bibinfo {author} {\bibfnamefont {C.-M.}\ \bibnamefont {Zhang}},\
  and\ \bibinfo {author} {\bibfnamefont {Q.}~\bibnamefont {Wang}},\ }\bibfield
  {title} {\bibinfo {title} {Twin-field quantum key distribution with modified
  coherent states},\ }\href@noop {} {\bibfield  {journal} {\bibinfo  {journal}
  {Opt. Lett.}\ }\textbf {\bibinfo {volume} {44}},\ \bibinfo {pages} {1468}
  (\bibinfo {year} {2019})}\BibitemShut {NoStop}%
\bibitem [{\citenamefont {Xu}\ \emph {et~al.}(2020{\natexlab{c}})\citenamefont
  {Xu}, \citenamefont {Hu}, \citenamefont {Feng},\ and\ \citenamefont
  {Wang}}]{xu2020hybrid}%
  \BibitemOpen
  \bibfield  {author} {\bibinfo {author} {\bibfnamefont {H.}~\bibnamefont
  {Xu}}, \bibinfo {author} {\bibfnamefont {X.-L.}\ \bibnamefont {Hu}}, \bibinfo
  {author} {\bibfnamefont {X.-L.}\ \bibnamefont {Feng}},\ and\ \bibinfo
  {author} {\bibfnamefont {X.-B.}\ \bibnamefont {Wang}},\ }\bibfield  {title}
  {\bibinfo {title} {Hybrid protocol for sending-or-not-sending twin-field
  quantum key distribution},\ }\href@noop {} {\bibfield  {journal} {\bibinfo
  {journal} {Optics Letters}\ }\textbf {\bibinfo {volume} {45}},\ \bibinfo
  {pages} {4120} (\bibinfo {year} {2020}{\natexlab{c}})}\BibitemShut {NoStop}%
\bibitem [{\citenamefont {Neergaard-Nielsen}\ \emph {et~al.}(2006)\citenamefont
  {Neergaard-Nielsen}, \citenamefont {Nielsen}, \citenamefont {Hettich},
  \citenamefont {M\o{}lmer},\ and\ \citenamefont
  {Polzik}}]{PhysRevLett.97.083604}%
  \BibitemOpen
  \bibfield  {author} {\bibinfo {author} {\bibfnamefont {J.~S.}\ \bibnamefont
  {Neergaard-Nielsen}}, \bibinfo {author} {\bibfnamefont {B.~M.}\ \bibnamefont
  {Nielsen}}, \bibinfo {author} {\bibfnamefont {C.}~\bibnamefont {Hettich}},
  \bibinfo {author} {\bibfnamefont {K.}~\bibnamefont {M\o{}lmer}},\ and\
  \bibinfo {author} {\bibfnamefont {E.~S.}\ \bibnamefont {Polzik}},\ }\bibfield
   {title} {\bibinfo {title} {Generation of a superposition of odd photon
  number states for quantum information networks},\ }\href
  {https://doi.org/10.1103/PhysRevLett.97.083604} {\bibfield  {journal}
  {\bibinfo  {journal} {Phys. Rev. Lett.}\ }\textbf {\bibinfo {volume} {97}},\
  \bibinfo {pages} {083604} (\bibinfo {year} {2006})}\BibitemShut {NoStop}%
\bibitem [{\citenamefont {Huang}\ \emph {et~al.}(2015)\citenamefont {Huang},
  \citenamefont {Le~Jeannic}, \citenamefont {Ruaudel}, \citenamefont {Verma},
  \citenamefont {Shaw}, \citenamefont {Marsili}, \citenamefont {Nam},
  \citenamefont {Wu}, \citenamefont {Zeng}, \citenamefont {Jeong},
  \citenamefont {Filip}, \citenamefont {Morin},\ and\ \citenamefont
  {Laurat}}]{PhysRevLett.115.023602}%
  \BibitemOpen
  \bibfield  {author} {\bibinfo {author} {\bibfnamefont {K.}~\bibnamefont
  {Huang}}, \bibinfo {author} {\bibfnamefont {H.}~\bibnamefont {Le~Jeannic}},
  \bibinfo {author} {\bibfnamefont {J.}~\bibnamefont {Ruaudel}}, \bibinfo
  {author} {\bibfnamefont {V.~B.}\ \bibnamefont {Verma}}, \bibinfo {author}
  {\bibfnamefont {M.~D.}\ \bibnamefont {Shaw}}, \bibinfo {author}
  {\bibfnamefont {F.}~\bibnamefont {Marsili}}, \bibinfo {author} {\bibfnamefont
  {S.~W.}\ \bibnamefont {Nam}}, \bibinfo {author} {\bibfnamefont
  {E.}~\bibnamefont {Wu}}, \bibinfo {author} {\bibfnamefont {H.}~\bibnamefont
  {Zeng}}, \bibinfo {author} {\bibfnamefont {Y.-C.}\ \bibnamefont {Jeong}},
  \bibinfo {author} {\bibfnamefont {R.}~\bibnamefont {Filip}}, \bibinfo
  {author} {\bibfnamefont {O.}~\bibnamefont {Morin}},\ and\ \bibinfo {author}
  {\bibfnamefont {J.}~\bibnamefont {Laurat}},\ }\bibfield  {title} {\bibinfo
  {title} {Optical synthesis of large-amplitude squeezed coherent-state
  superpositions with minimal resources},\ }\href
  {https://doi.org/10.1103/PhysRevLett.115.023602} {\bibfield  {journal}
  {\bibinfo  {journal} {Phys. Rev. Lett.}\ }\textbf {\bibinfo {volume} {115}},\
  \bibinfo {pages} {023602} (\bibinfo {year} {2015})}\BibitemShut {NoStop}%
\bibitem [{\citenamefont {Sychev}\ \emph {et~al.}(2017)\citenamefont {Sychev},
  \citenamefont {Ulanov}, \citenamefont {Pushkina}, \citenamefont {Richards},
  \citenamefont {Fedorov},\ and\ \citenamefont {Lvovsky}}]{2017Enlargement}%
  \BibitemOpen
  \bibfield  {author} {\bibinfo {author} {\bibfnamefont {D.~V.}\ \bibnamefont
  {Sychev}}, \bibinfo {author} {\bibfnamefont {A.~E.}\ \bibnamefont {Ulanov}},
  \bibinfo {author} {\bibfnamefont {A.~A.}\ \bibnamefont {Pushkina}}, \bibinfo
  {author} {\bibfnamefont {M.~W.}\ \bibnamefont {Richards}}, \bibinfo {author}
  {\bibfnamefont {I.~A.}\ \bibnamefont {Fedorov}},\ and\ \bibinfo {author}
  {\bibfnamefont {A.~I.}\ \bibnamefont {Lvovsky}},\ }\bibfield  {title}
  {\bibinfo {title} {Enlargement of optical schrdinger's cat states},\
  }\href@noop {} {\bibfield  {journal} {\bibinfo  {journal} {Nature Photonics}\
  }\textbf {\bibinfo {volume} {11}},\ \bibinfo {pages} {379} (\bibinfo {year}
  {2017})}\BibitemShut {NoStop}%
\bibitem [{\citenamefont {Cao}\ \emph {et~al.}(2015)\citenamefont {Cao},
  \citenamefont {Zhang}, \citenamefont {Lo},\ and\ \citenamefont
  {Ma}}]{Cao_2015}%
  \BibitemOpen
  \bibfield  {author} {\bibinfo {author} {\bibfnamefont {Z.}~\bibnamefont
  {Cao}}, \bibinfo {author} {\bibfnamefont {Z.}~\bibnamefont {Zhang}}, \bibinfo
  {author} {\bibfnamefont {H.-K.}\ \bibnamefont {Lo}},\ and\ \bibinfo {author}
  {\bibfnamefont {X.}~\bibnamefont {Ma}},\ }\bibfield  {title} {\bibinfo
  {title} {Discrete-phase-randomized coherent state source and its application
  in quantum key distribution},\ }\href
  {https://doi.org/10.1088/1367-2630/17/5/053014} {\bibfield  {journal}
  {\bibinfo  {journal} {New Journal of Physics}\ }\textbf {\bibinfo {volume}
  {17}},\ \bibinfo {pages} {053014} (\bibinfo {year} {2015})}\BibitemShut
  {NoStop}%
\bibitem [{\citenamefont {Chernoff}(1952)}]{chernoff1952measure}%
  \BibitemOpen
  \bibfield  {author} {\bibinfo {author} {\bibfnamefont {H.}~\bibnamefont
  {Chernoff}},\ }\bibfield  {title} {\bibinfo {title} {A measure of asymptotic
  efficiency for tests of a hypothesis based on the sum of observations},\
  }\href {http://www.jstor.org/stable/2236576} {\bibfield  {journal} {\bibinfo
  {journal} {Ann. Math. Stat.}\ }\textbf {\bibinfo {volume} {23}},\ \bibinfo
  {pages} {493} (\bibinfo {year} {1952})}\BibitemShut {NoStop}%
\bibitem [{\citenamefont {Yin}\ \emph {et~al.}(2020)\citenamefont {Yin},
  \citenamefont {Zhou}, \citenamefont {Gu}, \citenamefont {Xie}, \citenamefont
  {Lu},\ and\ \citenamefont {Chen}}]{yin2020tight}%
  \BibitemOpen
  \bibfield  {author} {\bibinfo {author} {\bibfnamefont {H.-L.}\ \bibnamefont
  {Yin}}, \bibinfo {author} {\bibfnamefont {M.-G.}\ \bibnamefont {Zhou}},
  \bibinfo {author} {\bibfnamefont {J.}~\bibnamefont {Gu}}, \bibinfo {author}
  {\bibfnamefont {Y.-M.}\ \bibnamefont {Xie}}, \bibinfo {author} {\bibfnamefont
  {Y.-S.}\ \bibnamefont {Lu}},\ and\ \bibinfo {author} {\bibfnamefont {Z.-B.}\
  \bibnamefont {Chen}},\ }\bibfield  {title} {\bibinfo {title} {Tight security
  bounds for decoy-state quantum key distribution},\ }\href@noop {} {\bibfield
  {journal} {\bibinfo  {journal} {Sci. Rep.}\ }\textbf {\bibinfo {volume}
  {10}},\ \bibinfo {pages} {14312} (\bibinfo {year} {2020})}\BibitemShut
  {NoStop}%
\bibitem [{\citenamefont {Yin}\ \emph {et~al.}(2023)\citenamefont {Yin},
  \citenamefont {Fu}, \citenamefont {Li}, \citenamefont {Weng}, \citenamefont
  {Li}, \citenamefont {Gu}, \citenamefont {Lu}, \citenamefont {Huang},\ and\
  \citenamefont {Chen}}]{yin2023experimental}%
  \BibitemOpen
  \bibfield  {author} {\bibinfo {author} {\bibfnamefont {H.-L.}\ \bibnamefont
  {Yin}}, \bibinfo {author} {\bibfnamefont {Y.}~\bibnamefont {Fu}}, \bibinfo
  {author} {\bibfnamefont {C.-L.}\ \bibnamefont {Li}}, \bibinfo {author}
  {\bibfnamefont {C.-X.}\ \bibnamefont {Weng}}, \bibinfo {author}
  {\bibfnamefont {B.-H.}\ \bibnamefont {Li}}, \bibinfo {author} {\bibfnamefont
  {J.}~\bibnamefont {Gu}}, \bibinfo {author} {\bibfnamefont {Y.-S.}\
  \bibnamefont {Lu}}, \bibinfo {author} {\bibfnamefont {S.}~\bibnamefont
  {Huang}},\ and\ \bibinfo {author} {\bibfnamefont {Z.-B.}\ \bibnamefont
  {Chen}},\ }\bibfield  {title} {\bibinfo {title} {Experimental quantum secure
  network with digital signatures and encryption},\ }\href@noop {} {\bibfield
  {journal} {\bibinfo  {journal} {Natl. Sci. Rev.}\ }\textbf {\bibinfo {volume}
  {10}},\ \bibinfo {pages} {nwac228} (\bibinfo {year} {2023})}\BibitemShut
  {NoStop}%
\bibitem [{\citenamefont {Yu}\ \emph {et~al.}(2015)\citenamefont {Yu},
  \citenamefont {Zhou},\ and\ \citenamefont {Wang}}]{yu2015statistical}%
  \BibitemOpen
  \bibfield  {author} {\bibinfo {author} {\bibfnamefont {Z.-W.}\ \bibnamefont
  {Yu}}, \bibinfo {author} {\bibfnamefont {Y.-H.}\ \bibnamefont {Zhou}},\ and\
  \bibinfo {author} {\bibfnamefont {X.-B.}\ \bibnamefont {Wang}},\ }\bibfield
  {title} {\bibinfo {title} {Statistical fluctuation analysis for
  measurement-device-independent quantum key distribution with three-intensity
  decoy-state method},\ }\href@noop {} {\bibfield  {journal} {\bibinfo
  {journal} {Phys. Rev. A}\ }\textbf {\bibinfo {volume} {91}},\ \bibinfo
  {pages} {032318} (\bibinfo {year} {2015})}\BibitemShut {NoStop}%
\end{thebibliography}

%

\end{document}